\pdfoutput=1
\documentclass[cits]{JINST}

\usepackage{amssymb}        
\usepackage[amssymb]{SIunits}
\usepackage{amsmath}

\usepackage{graphicx}
\usepackage{tikz}   
\usepackage{3dplot}  
\definecolor{dark-gray}{gray}{0.3}

\title{Calibrating the absolute amplitude scale for air showers measured at LOFAR}

\author{A.~Nelles$^{a,b}$\thanks{Corresponding author.}, J.~R.~H\"orandel$^{a,c}$, T.~Karskens$^a$, M.~Krause$^{a,d}$, S.~Buitink$^e$, A.~Corstanje$^a$, J.~E.~Enriquez$^a$, M.~Erdmann$^f$, H.~Falcke$^{a,c,g,h}$, A.~Haungs$^i$, R.~Hiller$^i$, T.~Huege$^i$, R.~Krause$^f$,  K.~Link$^i$, M.~J.~Norden$^g$, J.~P.~Rachen$^a$, L.~Rossetto$^a$, P.~Schellart$^a$, O.~Scholten$^j$, F.~G.~Schr\"oder$^i$, S.~ter Veen$^{a,g}$, S.~Thoudam$^a$, T.~N.~G.~Trinh$^j$, K.~Weidenhaupt$^f$, S.~J.~Wijnholds$^g$,
J.~Anderson$^k$, L.~B\"ahren$^l$, M.~E.~Bell$^m$, M.~J.~Bentum$^{g,n}$, P.~Best$^o$, A.~Bonafede$^p$, J.~Bregman$^g$, W.~N.~Brouw$^{g,q}$,
M.~Br\"uggen$^p$, H.~R.~Butcher$^r$, D.~Carbone$^l$, B.~Ciardi$^s$, F.~de Gasperin$^p$, S.~Duscha$^g$, J.~Eisl\"offel$^t$,
R.~A.~Fallows$^g$, W.~Frieswijk$^g$, M.~A.~Garrett$^{g,u}$, M.~P.~van Haarlem$^g$, G.~Heald$^{g,q}$, M.~Hoeft$^t$, 
A.~Horneffer$^h$, M.~Iacobelli$^g$, E.~Juette$^v$, A.~Karastergiou$^w$,
J.~Kohler$^h$, V.~I.~Kondratiev$^{g,x}$, M.~Kuniyoshi$^y$,  G.~Kuper$^g$,
J.~van Leeuwen$^{g,l}$, P.~Maat$^g$, R.~McFadden$^g$, D.~McKay-Bukowski$^{z,aa}$,
E.~Orru$^g$, H.~Paas$^{ab}$, M.~Pandey-Pommier$^{ac}$,  V.~N.~Pandey$^g$, 
R.~Pizzo$^g$, A.~G.~Polatidis$^g$, W.~Reich$^h$, H.~R\"ottgering$^u$,  D.~Schwarz$^{ad}$, M.~Serylak$^w$,  J.~Sluman$^g$, 
O.~Smirnov$^{ae,af}$,  C.~Tasse$^{ag}$,  M.~C.~Toribio$^g$,   R.~Vermeulen$^g$, R.~J.~van Weeren$^{ah}$,  R.~A.~M.~J.~Wijers$^l$,
O.~Wucknitz$^h$, P.~Zarka$^{ag}$\\
\llap{$^a$}Department of Astrophysics/IMAPP, Radboud University Nijmegen, P.O. Box 9010, 6500 GL Nijmegen, The Netherlands\\
\llap{$^b$}now at Department of Physics and Astronomy, University of California Irvine, Irvine, CA, 92697, USA\\
\llap{$^c$}Nikhef, Science Park Amsterdam, 1098 XG Amsterdam, The Netherlands\\
\llap{$^d$}now at DESY, Platanenallee 6, 15738 Zeuthen, Germany\\
\llap{$^e$}Astrophysical Institute, Vrije Universiteit Brussel, Pleinlaan 2, 1050 Brussels, Belgium\\
\llap{$^f$}RWTH Aachen University, III. Physikalisches Institut A, 52056 Aachen, Germany\\
\llap{$^g$}ASTRON, the Netherlands Institute for Radio Astronomy, Postbus 2, 7990 AA Dwingeloo, The Netherlands\\
\llap{$^h$}Max-Planck-Institut f\"ur Radioastronomie, Auf dem H\"ugel 69, 53121 Bonn, Germany\\
\llap{$^i$}Institut f\"ur Kernphysik, Karlsruhe Institute of Technology (KIT), Hermann-von-Helmholtz-Platz 1, 76344 Eggenstein-Leopoldshafen, Germany\\
\llap{$^j$}KVI-CART, University Groningen, P.O. Box 72, 9700 AB Groningen, The Netherlands\\
\llap{$^k$}Helmholtz-Zentrum Potsdam, DeutschesGeoForschungsZentrum GFZ, Department 1: Geodesy and Remote Sensing, Telegrafenberg, A17, 14473 Potsdam, Germany\\
\llap{$^l$}Anton Pannekoek Institute, University of Amsterdam, Postbus 94249, 1090 GE Amsterdam, The Netherlands \\
\llap{$^m$}CSIRO Australia Telescope National Facility, PO Box 76, Epping NSW 1710, Australia\\
\llap{$^n$}University of Twente, The Netherlands\\
\llap{$^o$}Institute for Astronomy, University of Edinburgh, Royal Observatory of Edinburgh, Blackford Hill, Edinburgh EH9 3HJ, UK \\
\llap{$^p$}University of Hamburg, Gojenbergsweg 112, 21029 Hamburg, Germany\\
\llap{$^q$}Kapteyn Astronomical Institute, PO Box 800, 9700 AV Groningen, The Netherlands \\
\llap{$^r$}Research School of Astronomy and Astrophysics, Australian National University, Mt Stromlo Obs., via Cotter Road, Weston, A.C.T. 2611, Australia\\
\llap{$^s$}Max Planck Institute for Astrophysics, Karl Schwarzschild Str. 1, 85741 Garching, Germany \\
\llap{$^t$}Th\"{u}ringer Landessternwarte, Sternwarte 5, D-07778 Tautenburg, Germany\\
\llap{$^u$}Leiden Observatory, Leiden University, PO Box 9513, 2300 RA Leiden, The Netherlands \\
\llap{$^v$}Astronomisches Institut der Ruhr-Universit\"{a}t Bochum, Universitaetsstrasse 150, 44780 Bochum, Germany \\
\llap{$^w$}Astrophysics, University of Oxford, Denys Wilkinson Building, Keble Road, Oxford OX1 3RH, UK \\
\llap{$^x$}Astro Space Center of the Lebedev Physical Institute, Profsoyuznaya str. 84/32, Moscow 117997, Russia \\
\llap{$^y$}National Astronomical Observatory of Japan, Japan \\
\llap{$^{z}$}Sodankyl\"{a} Geophysical Observatory, University of Oulu, T\"{a}htel\"{a}ntie 62, 99600 Sodankyl\"{a}, Finland\\
\llap{$^{aa}$}STFC Rutherford Appleton Laboratory,  Harwell Science and Innovation Campus,  Didcot  OX11 0QX, UK \\
\llap{$^{ab}$}Center for Information Technology (CIT), University of Groningen, The Netherlands \\
\llap{$^{ac}$}Centre de Recherche Astrophysique de Lyon, Observatoire de Lyon, 9 av Charles Andr\'{e}, 69561 Saint Genis Laval Cedex, France \\
\llap{$^{ad}$} Fakult\"{a}t f\"{u}r Physik, Universit\"{a}t Bielefeld, Postfach 100131, D-33501, Bielefeld, Germany \\
\llap{$^{ae}$}Department of Physics and Elelctronics, Rhodes University, PO Box 94, Grahamstown 6140, South Africa \\
\llap{$^{af}$}SKA South Africa, 3rd Floor, The Park, Park Road, Pinelands, 7405, South Africa \\
\llap{$^{ag}$}LESIA, UMR CNRS 8109, Observatoire de Paris, 92195   Meudon, France \\
\llap{$^{ah}$}Harvard-Smithsonian Center for Astrophysics, 60 Garden Street, Cambridge, MA 02138, USA \\

E-mail: \email{a.nelles@astro.ru.nl}}

\abstract{Air showers induced by cosmic rays create nanosecond pulses detectable at radio frequencies. These pulses have been measured successfully in the past few years at the LOw-Frequency ARray (LOFAR) and are used to study the properties of cosmic rays. For a complete understanding of this phenomenon and the underlying physical processes, an absolute calibration of the detecting antenna system is needed. We present three approaches that were used to check and improve the antenna model of LOFAR and to provide an absolute calibration of the whole system for air shower measurements. Two methods are based on calibrated reference sources and one on a calibration approach using the diffuse radio emission of the Galaxy, optimized for short data-sets. An accuracy of 19\% in amplitude is reached. The absolute calibration is also compared to predictions from air shower simulations. These results are used to set an absolute energy scale for air shower measurements and can be used as a basis for an absolute scale for the measurement of astronomical transients with LOFAR. }


\begin{document}
\section{Introduction}
When a high-energy cosmic ray hits the top of the Earth's atmosphere, a cascade of secondary particles is created that propagates through the atmosphere to the ground. This phenomenon is called an extensive air shower.  A promising method for the detection of air showers is based on their radio emission. Air showers emit short broad-band pulses that are caused by the electromagnetic component of the showers \cite{Allan1971}. As the shower develops in the magnetic field of the Earth, a time-varying transverse current of electrons and positrons is created, which causes radio emission \cite{KahnLerche1966}. Also, air showers accumulate electrons, resulting in a charge excess in the shower front. This again induces a time-varying current, which leads to emission \cite{Askaryan1965}. The emission from both currents adds up coherently in frequencies below $\unit{200}{MHz}$ that correspond to the dimensions of the shower front, i.e.~a couple of meters \cite{Huege2013b,Alvarez-Muniz2011,Werner2012}. Measuring the radio emission of air showers has shown to be a good method to determine the type and the energy of the primary particle \cite{Buitink2014}, \cite{Apel2014}.

The LOw-Frequency ARray (LOFAR) is used to study the radio emission processes of air showers \cite{vanHaarlem2013,Schellart2013,Nelles2015a}. An understanding of the electric field of the radio pulse in each antenna is essential for the reconstruction of the air showers. For this, the response of the electronic system and antenna to the incoming signal is most important. The recorded voltage traces depend severely on the instrumental characteristics, such as the antenna gain, filter amplifiers or losses in cables. In addition, the gain-patterns of the used antennas are direction dependent, which has to be considered for the different arrival directions of the cosmic rays. In order to fully understand the phenomenon of radio emission and the underlying physical processes, an absolute calibration of the measured field strength is needed. This also enables us to test the absolute energy scale of different experiments such as LOPES \cite{Falcke:2005aa}, Tunka-Rex \cite{Kostunin:2013}, or the Auger Engineering Radio Array (AERA) \cite{Schroeder:2013b}, \cite{Aab2014} of the Pierre Auger Observatory \cite{Auger}. Such tests are independent of air shower simulations and allow us to compare the results of the calibrations conducted by different groups. 

For astronomical observations of sources, essentially at infinite distances, a number of calibration procedures are used by the LOFAR Collaboration. They mostly depend on known point-like astronomical sources which are resolved in imaging and used for flux calibration. Thereby, they are based on integrated or correlated quantities, requiring a significant amount of data. The signals of air showers are non-repeating pulses of a duration of less than $\unit{100}{ns}$ and usually just $\unit{2.1}{\mu s}$ of data are recorded per antenna and measurement. Therefore, a calibration based on resolved point-sources is challenging with our data.  

We present the results of two measurement campaigns that were conducted at LOFAR in order to understand and calibrate the low-band antenna together with the LOFAR system (Section \ref{sec:lofar} and \ref{sec:set-ups}). Both include reference sources that were placed in the vicinity of and measured with LOFAR. One method is used to verify the antenna model with respect to the directional as well as frequency dependent response of the antenna (Section \ref{sec:veri}). The other is used to provide an absolute calibration of the system. This is accompanied by a third method that is based on calculations using all-sky emission models of the Galaxy (Section \ref{sec:Abs}). The results are also compared to those obtained with the same method at other experiments \cite{AugerAntennas2012,Nehls2008,TunkaCal}. For reference purposes also a comparison to the absolute field strength as predicted by air shower simulations is shown in Section \ref{sec:coreas}.


\begin{figure}
\centering
\includegraphics[width=0.43\textwidth]{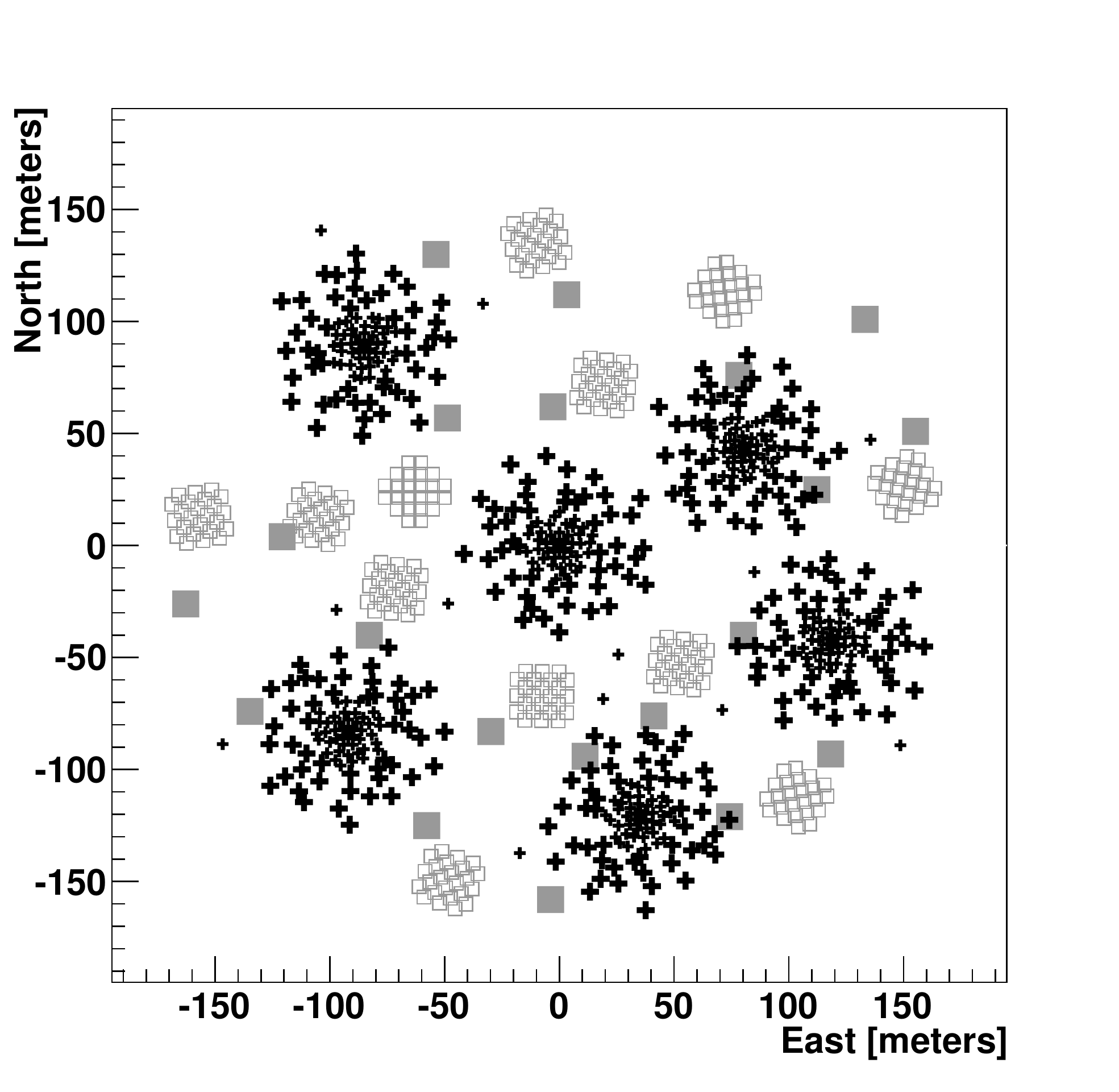}
 \includegraphics[width=0.52\textwidth]{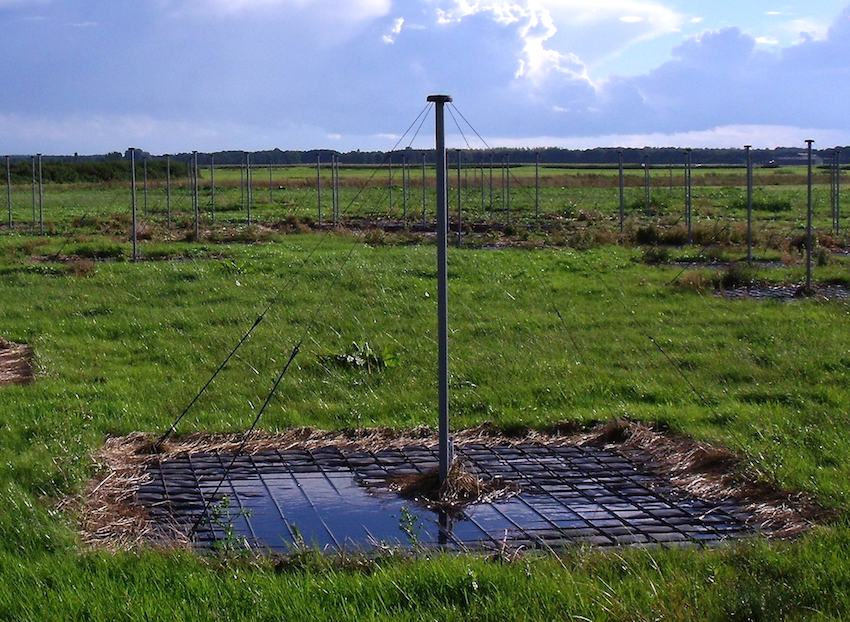}
\caption{Left: The six innermost stations of the Superterp. The black crosses indicate the positions of the LBA antennas, the gray full squares indicate the LORA scintillation detectors and the open squares mark the HBA tiles. One circular cluster of LBA antennas  and two clusters of HBA tiles belong to every station. Right: A low-band antenna at LOFAR. The supporting vertical PVC pipe is shown with the LNA on top of it. The dipole wires are located to the right and left of the LBA and are attached to the wired ground-plate }
\label{fig:LBA}
\end{figure}

\section{The LOFAR antennas}
\label{sec:lofar}
LOFAR antenna stations are distributed over the Netherlands and Europe \cite{vanHaarlem2013}, \cite{LOFARADD}. Nine international and thirty-eight Dutch stations make up an interferometric array of dipole antennas with effectively all-sky coverage. The core of LOFAR, located in the North of the Netherlands, consists of 24 antenna stations within an area of \unit{12}{km^{2}}. The \emph{Superterp} consists of the six innermost stations inside the core within a diameter of \unit{320}{m} as shown in Figure \ref{fig:LBA}.

A LOFAR station comprises two types of antennas, the Low-Band Antennas (LBAs) \cite{LBAADD}, as well as the High-Band Antennas (HBAs) \cite{HBAADD}. Together, they cover the frequency range from 10 to \unit{240}{\mega Hz}. Every core station in the Netherlands consists of 96 LBAs, 48 HBAs and 96 digital receiver units (RCUs) \cite{vanHaarlem2013}. The LBAs are grouped into an inner and outer circle that each contain 48 dipoles and are called LBA Inner and LBA Outer, respectively. All read-out electronics are equipped with RAM buffers on transient buffer boards (TBBs) that store up to \unit{5}{s} of data. These buffers can be read out after a trigger or on manual request. At the Superterp, an array of 20 particle detectors is used to trigger a read-out of the LOFAR TBBs whenever an air shower above a minimum energy threshold has been detected \cite{Thoudam2013}. The data are then available as individual time-domain signals per antenna and are combined in one file for every station.

\subsection{The Low-Band Antennas}
\label{sec:LBA}
The LBAs are designed to measure in the frequency range between 10 and \unit{90}{\mega Hz} \cite{vanHaarlem2013}, \cite{LBAADD}, where the lower end is determined by the strong ionospheric reflection of terrestrial radio-frequency interference (RFI), and where the higher end is marked by the commercial FM radio band. Practically, the operational range is therefore limited to \unit{30-80}{\mega Hz}.

An LBA in the field of LOFAR is shown in Figure \ref{fig:LBA}. The sensitive elements of the antenna are four copper wires that act as slanted dipoles. The antenna is held upright with a PVC pipe and with tension on the springs in which the copper wires terminate. Two coaxial cables that carry the signal of the two polarizations run through the PVC pipe. These cables are also used for the power supply of the low-noise amplifier (LNA) that is on top of the LBA and acts as a voltage amplifier, including a notch filter for the FM band \cite{LBAADD}. According to the length of the wires, the resonance frequency of the LBA dipole should be \unit{58}{\mega Hz}. However, this does not include the influences of the other electronics. A different resonance frequency is obtained when including the whole analogue and digital system in the model. The antennas are known to be sky-noise dominated. Observations of CAS A and 3C295 were used to confirm that the fraction of sky-noise over total noise is above 0.5 for most of the LBA frequency range \cite{vanHaarlem2013}, \cite{Wijnholds2011}.

Every LBA has two polarizations. One polarization of the antenna is aligned from northeast to southwest, while the other is aligned orthogonally from northwest to southeast. The omnidirectional response, which is provided by the two orthogonal polarizations, allows for simultaneous observations of the whole visible sky. 

\subsection{System characteristics and antenna simulation}
In order to obtain an absolute calibration for the LOFAR system, a frequency and direction dependent factor is needed that converts the measured voltage traces (in uncalibrated units) to the electric field emitted from the incoming air shower. Two contributions have to be considered: the vector effective length of the antenna (VEL), and the additional frequency dependent gain and phase factors introduced by filters and cables.\footnote{This section summarizes briefly some aspects of antenna theory and introduces our notation. More detailed descriptions can be found in e.g. \cite{AugerAntennas2012,LoLee1993,Balanis2005}} The VEL contains all directional sensitivity, as well as a large part of the frequency dependent gain and group delay. Understanding the antenna will therefore allow us to compare signals of air showers arriving from different directions. In order to achieve this, the entire system has to be considered. 

Every incoming electric field signal $\vec{E}(t)$ can be separated into its two components perpendicular to the direction of propagation, indicating its two polarization components in $\hat{e}_{\theta}$ and $\hat{e}_{\phi}$:
\begin{equation}
\vec{E}(t) = \vec{E}_{\theta}(t)\hat{e}_{\theta} +  \vec{E}_{\phi}(t)\hat{e}_{\phi}.
\end{equation}
For the influence of the VEL of the antenna $\vec{H}$ that converts these electric fields $\vec{E}$ to voltages $V$, one can also consider them in the frequency domain rather than in the time-domain\footnote{For this analysis the focus is on the amplitude response. Future work will also include the phase response of the system to transient signals.}
\begin{equation}
V (\nu,\theta,\phi) = \vec{H}(\nu,\theta,\phi) \cdot \vec{E}(\nu,\theta,\phi),
\end{equation}
where $\theta$ denotes the zenith angle and $\phi$ the azimuth angle as indicated in Figure \ref{fig: COsystem} and $\nu$ is the frequency. 

\tdplotsetmaincoords{60}{120}
\pgfmathsetmacro{\rvec}{0.85}
\pgfmathsetmacro{\thetavec}{30}
\pgfmathsetmacro{\phivec}{60}
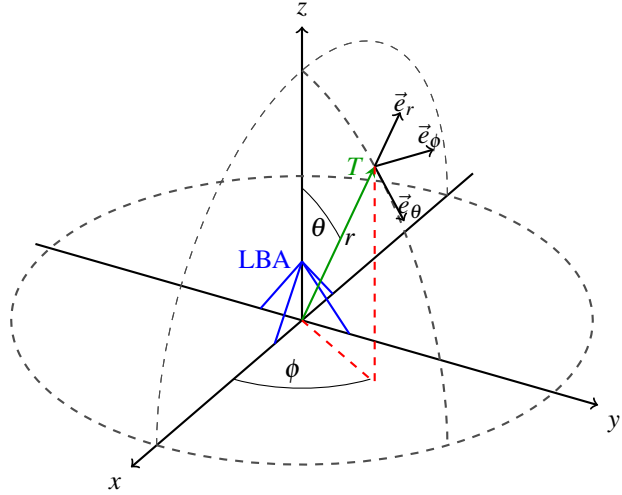
\begin{figure}
 \centering
\begin{tikzpicture}[scale=4.5,tdplot_main_coords]
\small
\coordinate (O) at (0,0,0);
\tdplotsetcoord{T}{\rvec}{\thetavec}{\phivec}
\draw[thick,->] (-1,0,0) -- (1,0,0) node[anchor=north east]{$x$};
\draw[thick,->] (0,-0.9,0) -- (0,1,0) node[anchor=north west]{$y$};
\draw[thick,->] (0,0,0) -- (0,0,1) node[anchor=south]{$z$};
\draw[thick,color=blue] (0,0,0.2) -- (0,-0.14,0);
\draw[thick,color=blue] (0,0,0.2) -- (0,0.16,0);
\draw[thick,color=blue] (0,0,0.2) -- (0.16,0,0);
\draw[thick,color=blue] (0,0,0.2) -- (-0.18,0,0);
\tdplotdrawarc{(O)}{0.4}{0}{\phivec}{}{}
\node at (25:.3){$\phi$};
\tdplotsetthetaplanecoords{\phivec}
\tdplotdrawarc[tdplot_rotated_coords]{(0,0,0)}{0.45}{0}{\thetavec}{anchor= north}{}
\node[tdplot_rotated_coords] at (15:0.38){$\theta$};
\draw[thick, dashed,tdplot_rotated_coords,color=dark-gray] (\rvec,0,0) arc(0:90:\rvec);
\draw[thick, dashed,color=dark-gray] (\rvec,0,0) arc (0:360:\rvec);
\node[tdplot_rotated_coords] at (42:0.95){$\vec{e}_{\theta}$};
\node[tdplot_rotated_coords] at (37:1.25){$\vec{e}_{\phi}$};
\node[tdplot_rotated_coords] at (30:1.2){$\vec{e}_{r}$};
\node[tdplot_rotated_coords] at (34:0.5){$r$};
\node[tdplot_rotated_coords] at (-65:0.24){\textcolor{blue}{LBA}};
\tdplotsetrotatedcoords{\phivec}{\thetavec}{0}
\tdplotsetrotatedcoordsorigin{(T)}
\draw[thick,tdplot_rotated_coords,->] (0,0,0) -- (.2,0,0);
\draw[thick,tdplot_rotated_coords,->] (0,0,0) -- (0,.2,0);
\draw[thick,tdplot_rotated_coords,->] (0,0,0) -- (0,0,.3);
\draw[thick, -stealth,color=green!60!black] (O) -- (T) node[anchor=east]{$T$};
\draw[thick, dashed, color=red] (O) -- (Txy);
\draw[thick, dashed, color=red] (T) -- (Txy);
\foreach \angle in {0,0}
{
\coordinate (P) at (0,0,\sintheta);
\tdplotsetthetaplanecoords{\angle}
\tdplotdrawarc[dashed,tdplot_rotated_coords,color=dark-gray]{(O)}{\rvec}{-90}{90}{}{}
}
\end{tikzpicture}
\caption[Spherical reference coordinate system of an antenna.]{Spherical reference coordinate system of a low-band antenna of LOFAR. Point \textit{T} indicates the position of the transmitting source which is used for the gain calibration measurements of the LBA. The distance between transmitting and receiving antenna is denoted by $r$. Symbols $\theta$ and $\phi$ indicate the zenith angle and azimuth angle, respectively.}
\label{fig: COsystem}
\end{figure}

For each frequency the VEL can be described by the Jones matrix $J$ with its components for the $X$- and the $Y$-dipoles of the LOFAR LBA
\begin{equation}
\begin{pmatrix}
V_{X} & V_{Y}
\end{pmatrix}
=
\begin{pmatrix}
J_{X \theta} & J_{X\phi}\\
J_{Y \theta} & J_{Y\phi}
\end{pmatrix}
\cdot 
\begin{pmatrix}
E_{\theta} \\
E_{\phi}
\end{pmatrix}.
\label{eq:jones}
\end{equation}
Since both dipoles are independent, the equation can be separated into the two components. For LOFAR, the Jones matrix of the LBA is modeled with the electromagnetic simulation software WIPL-D \cite{Kolundzija2011} and a customized software model of the gains of the electronics. This model contains some simplifications, but describes the system to our current best knowledge. 

\begin{figure}
\includegraphics[width=0.5\textwidth]{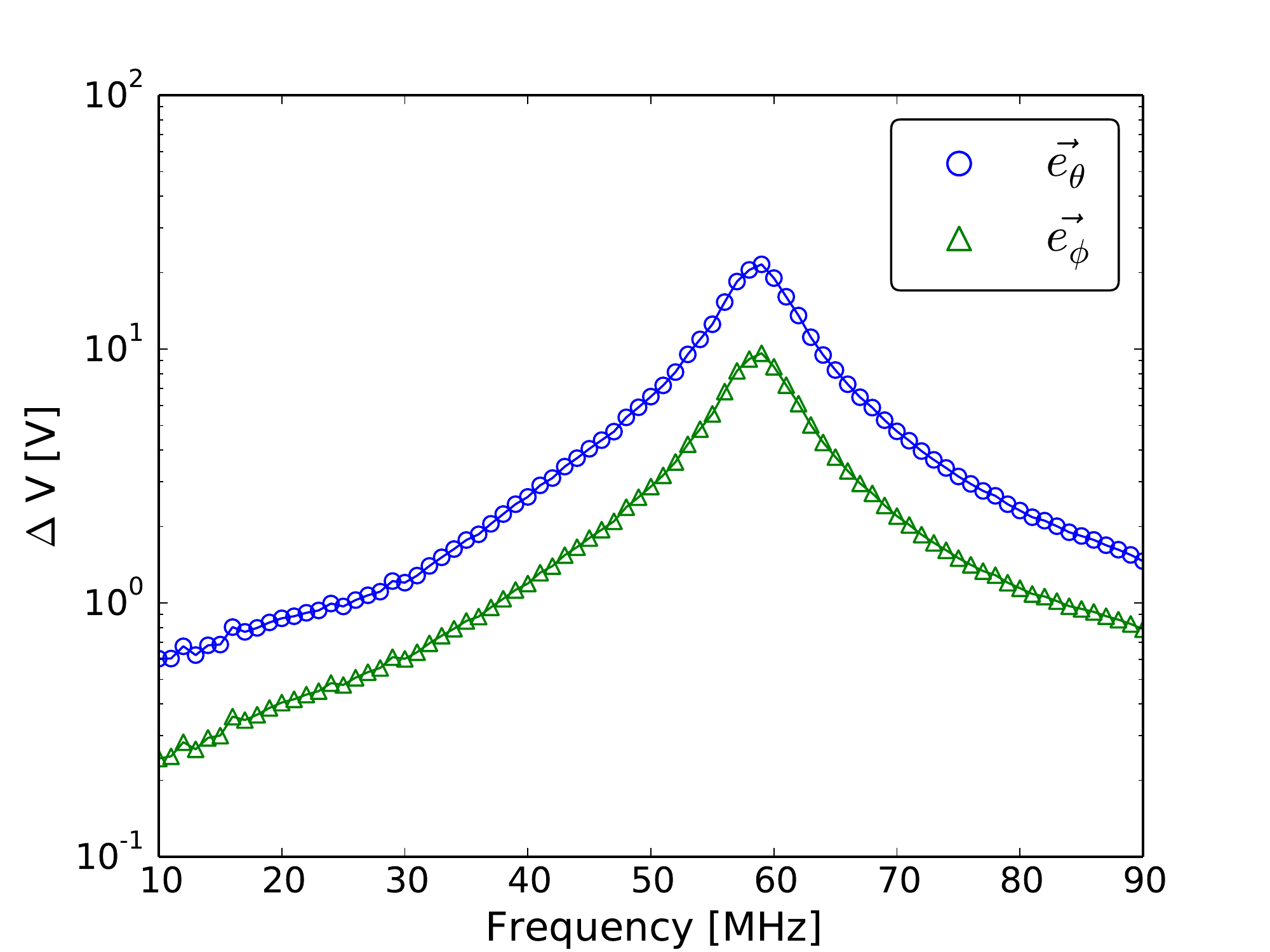}
\includegraphics[width=0.5\textwidth]{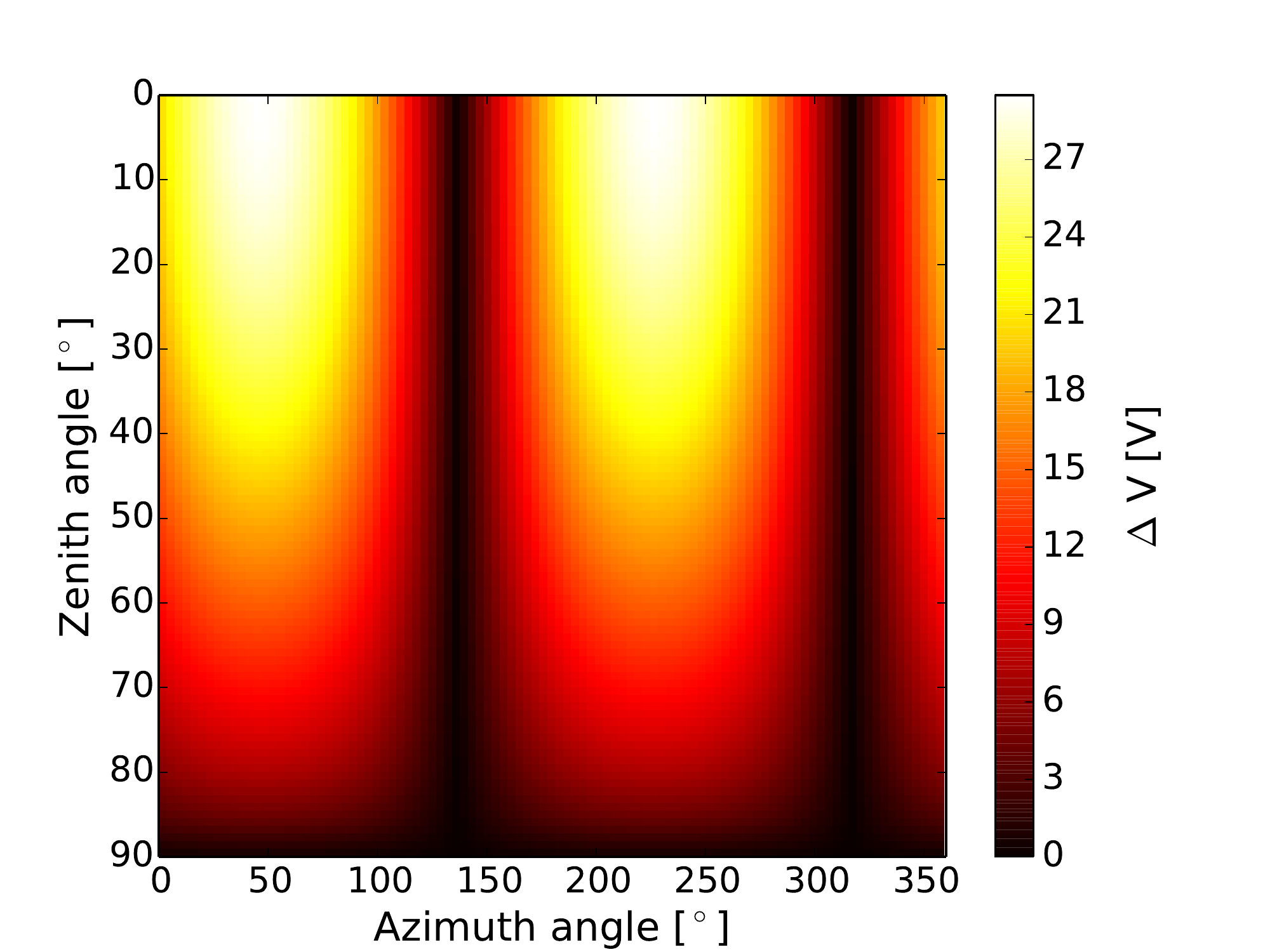}
\caption{Example response of the Vector Effective Length, in the form of output voltage $\Delta V$ for an LBA. Left: Response to an incoming wave polarized in the $\vec{e_\theta}$ direction (circles), and a wave polarized in the $\vec{e_\phi}$ direction (triangles), with arrival direction $\theta = 45^\circ$ and $\phi =  65^\circ$. Right: $|J_{X\theta}|$ component at $\unit{60}{MHz}$ as a function of direction for a wave fully polarized in the $\vec{e_\theta}$ direction ($E_{\theta}=1$).}
\label{fig:jones}
\end{figure}

The model predicts the response to an incoming wave with an electric field strength of \unit{1}{V/m}, and is computed with steps of \unit{1}{MHz} in frequency, \unit{5}{^\circ} in $\theta$ and \unit{10}{^\circ} in $\phi$. The components at intermediate values are obtained via trilinear interpolation when needed. An example of the directional sensitivity of the Jones matrix, as well as the frequency dependence are shown in Figure \ref{fig:jones}.

We can visualize the antenna system of the LBA together with the load impedance of the LNA in an equivalent circuit which is illustrated in Figure \ref{fig: circuit}. $V_{emf}$ represents the incoming signal, the voltage at the inputs of the LNA is denoted by $V_{in}$, the antenna impedance by $Z_{a}$ and $G_{V}$ is the voltage gain. As long as the input impedance $Z_{in}$ is large compared to the antenna impedance $Z_{a}$, there will be no voltage drop across the antenna impedance. The ratio of the output and the input voltage can be calculated by
\begin{equation}
 \frac{V_{\mathrm{out}}}{V_{\mathrm{emf}}}=G_{V}\cdot\frac{Z_{\mathrm{in}}}{Z_{\mathrm{in}}+Z_{a}}\cdot\frac{R_{\mathrm{out}}}{R_{\mathrm{out}}+Z_{\mathrm{out}}}
 \label{eq:equivalent circuit} 
\end{equation}
with 
\begin{equation}
 Z_{\mathrm{in}}=\frac{R}{1+j\omega RC},
\end{equation}
and $\omega=2\pi \nu$. The load impedance of the LNA $Z_{out}$ and the output resistance $R_{out}$ are \unit{75}{\ohm} for the LBA, respectively. This is also the typical resistor of the coaxial cables used in the LOFAR stations. Hence, the losses due to reflection are minimized. 

\definecolor{dark-gray}{gray}{0.3}
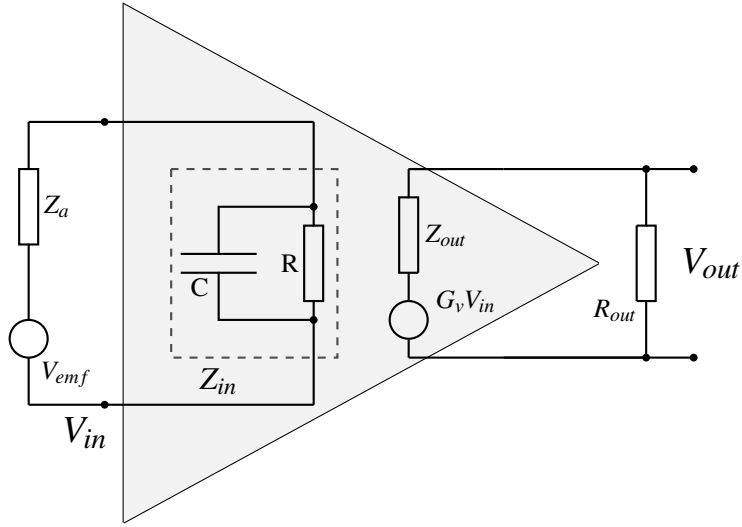
\begin{figure}[t]
 \centering
\begin{tikzpicture}[scale=1.25]
\small
\coordinate (O) at (0,0);
\draw[thick,color=black] (-3,-2.75) -- (-3,2.75)-- (2,0);
\draw[thick,color=black] (-3,-2.75) -- (2,0);
\fill [draw=none, fill=gray!10!white]  (-3,-2.75) -- (-3,2.75)-- (2,0)-- cycle;

\draw[thick,-] (0,-1) -- (0,-0.8);
\draw[thick,-] (0,-0.6) circle (0.2);
\draw[thick,-] (0,-0.4) -- (0,-0.1);
\draw[thick,-] (0,1) -- (0.4,1) ;
\node at (0.6,-0.4) {$G_{v}V_{in}$};

\draw[thick,-] (0,1) -- (0,0.7) -- (0.1,0.7)-- (0.1,-0.1);
\draw[thick,-] (0,0.7) -- (-0.1,0.7) -- (-0.1,-0.1) -- (0,-0.1) -- (0.1,-0.1);
\node at (0.4,0.3) {$Z_{out}$};

\draw[thick,-] (0,1.0) -- (1.0,1.0);
\draw[thick,fill=black] (2.5,1.0) circle (0.03);
\draw[thick,fill=black] (2.5,-1.0) circle (0.03);
\draw[thick,-] (2.5,1.0) -- (2.5,0.4) -- (2.4,0.4);
\draw[thick,-] (2.5,0.4) -- (2.6,0.4) -- (2.6,-0.4)  -- (2.5,-0.4);
\draw[thick,-] (2.4,0.4) -- (2.4,-0.4); 
\node at (2.18,-0.5) {$R_{out}$};
\draw[thick,-] (2.4,-0.4) -- (2.5,-0.4) -- (2.5,-1) -- (0,-1.0);
\draw[thick,-] (1,-1) -- (3,-1);
\draw[thick,-] (1,1) -- (3,1);
\draw[thick,fill=black] (3,1.0) circle (0.03);
\draw[thick,fill=black] (3,-1.0) circle (0.03);

\draw[thick,-]  (-1,-1.5) -- (-4,-1.5);
\draw[thick,-]  (-1,-1.5) -- (-1,-0.4) -- (-1.1,-0.4) -- (-1.1,0.4)-- (-1,0.4);
\node at (-1.25,0) {R};
\draw[thick,-]  (-1,-0.4) -- (-0.9,-0.4) -- (-0.9,0.4) -- (-1.,0.4)-- (-1,1.5);
\draw[thick,fill=black] (-1,0.6) circle (0.03);
\draw[thick,fill=black] (-1,-0.6) circle (0.03);
\draw[thick,-]  (-1,-0.6) -- (-2,-0.6) -- (-2,-0.1)-- (-2.4,-0.1);
\draw[thick,-]  (-2,0.1) -- (-2,0.6) -- (-1,0.6);
\draw[thick,-]  (-2.4,0.1) -- (-2,0.1) -- (-1.6,0.1);
\node at (-2.2,-0.25) {C};
\draw[thick,-]  (-2,-0.1) -- (-1.6,-0.1);
\draw[thick,-]  (-4,1.5) -- (-1,1.5);

\draw[thick,color=dark-gray,dashed]  (-2.5,1) -- (-2.5,-1)-- (-0.75,-1) -- (-0.75,1) -- (-2.5,1);
\node at (-2,-1.25) {\large $Z_{in}$};

\draw[thick,-]  (-4,1.5) -- (-4,1) -- (-4.1,1)-- (-4.1,0.2)-- (-4,0.2);
\draw[thick,-]  (-4,1) -- (-3.9,1) -- (-3.9,0.2)-- (-4.,0.2) -- (-4,-0.6);
\node at (-3.7,0.6) {$Z_{a}$};

\draw[thick,-] (-4,-0.8) circle (0.2);
\draw[thick,-]  (-4,-1) -- (-4,-1.5);
\node at (-3.6,-1.18) {$V_{emf}$};

\draw[thick,fill=black] (-3.2,-1.5) circle (0.03);
\draw[thick,fill=black] (-3.2,1.5) circle (0.03);
\node at (3.2,0) {\Large $V_{out}$};
\node at (-3.4,-1.8) {\Large $V_{in}$};
\end{tikzpicture}
\caption[Equivalent circuit of the LBA.]{Equivalent circuit of the LBA. The circuit represents a voltage source including an internal resistance which is equal to the antenna impedance. The LNA, which works as an operational amplifier, is located directly behind the antenna. The equivalent circuit consists of a capacitor $C$ and a resistor $R$. The output resistor $R_{out}$ has to be matched to the impedance of the LNA, $Z_{out}$. This matched situation corresponds to the calibration as performed with the flying reference source.}
\label{fig: circuit}
\end{figure}

 
\section{Calibration set-ups}
\label{sec:set-ups}
For the analysis of all measurements and to describe the radiation pattern of an antenna, a spherical coordinate system is used as shown in Figure \ref{fig: COsystem}. The origin of the coordinate system is located below the center of the antenna and all quantities refer to the electromagnetic far-field region. The emitted signal coming from large distances can be described as a plane wave. The vector of the electromagnetic field has an $\hat{e}_{\theta}$ and $\hat{e}_{\phi}$ component in the coordinate system of the transmitting antenna. The signal propagates in the $\hat{e}_{r}$ direction. 

It has been checked that all measurements have been conducted in the far-field region. The required minimum distance arises from far-field condition within the LBA operation bandwidth from \unit{30-80}{\mega Hz} and given the antenna dimensions $D=\unit{2.76}{m}$ to
\begin{equation}
 r\geq\frac{2D^{2}}{\lambda}\approx\unit{4}{m}.
\end{equation}
Consequently, with measurements taken at a distance of at least 12 m away from the source, the far-field approximations are valid for the measurements. However, the far-field criterion also includes the condition that $\lambda << r$ to avoid spill-over from the near-field into the far-field regime. For this condition the distances might be on the small side. It has been shown before that the spill-over effects were not observed for this particular experimental set-up \cite{Nehls2008}, while they were still relevant for other measurements \cite{AugerAntennas2012}. In this analysis they will be treated as additional uncertainties. 

In the two experimental set-ups, reference transmitters are used to calibrate the absolute scale of the LBAs. The emitted signals are significantly stronger than any contribution from system or ambient noise. Still, the signal strength was carefully adapted to not saturate the system and stay well below the non-linear regime of the amplifiers. All measurements were performed using regular LBAs installed in the LOFAR core. 

\definecolor{dark-gray}{gray}{0.4}
\begin{figure}
 \centering
\begin{tikzpicture}[scale=1.4]
\small
\coordinate (O) at (0,0);
\draw[thick, dotted,color=dark-gray] (-5.5,1.2) -- (-3.4,0.43) -- (-7.6,1.97);		
\draw[dashed,color=dark-gray] (-3.5,2.2) -- (-5.5,1.2);					
\draw[thick,color=black] (-5.5,1.2) -- (-5.5,-0.7) -- (-3.5,-0.7);
\draw[thick,color=black] (-3.5,-0.7) -- (-3.5,-0.5) -- (-2.5,-0.5) -- (-2.5,-0.9) -- (-3.5,-0.9) -- (-3.5,-0.7);
\draw[dashdotted, color=black] (-3.5,2.2) -- (-3.5,1.2) -- (-5.5,1.2);
\draw[thick,->,color=black] (-2.8,1.3) -- (-3.45,1.3);
\node[text width=3cm] at (-1.85,1.3) {Calibration Loop of Network Ana\-lyser};
\draw[thick,line width = 0.5mm, color=black] (-1.8,-2.0) -- (-1.8,-1.4) -- (-0.2,-1.4) -- (-0.2,-2.0) -- (-1.8,-2.0); 
\node at (-1.0,-2.2) {Vector Network Analyser};
\draw[thick,color=black] (-2.5,-0.7) -- (-1.5,-0.7) -- (-1.5,-1.7);			
\draw[thick,fill=green] (-1.5,-1.7) circle (0.08);
\draw[thick,color=black] (-0.5,-1.7) -- (-0.5,2.2) -- (-1.5,2.2);
\draw[thick,fill=green] (-0.5,-1.7) circle (0.08);
\draw[thick,color=black] (-1.5,2.2) -- (-1.5,2.0) -- (-2.5,2.0) -- (-2.5,2.4) -- (-1.5,2.4) -- (-1.5,2.2); 
\draw[thick,color=black] (-2.5,2.2) -- (-3.5,2.2);
\draw[thick,color=dark-gray] (-3.4,0.43) arc (-20:160:2.24);				

\draw[dashed,color=dark-gray] (-5.5,1.8) -- (-5.5,3.6);
\tdplotdrawarc[color=gray,thick]{(-5.5,1.2)}{1.0}{90}{26}{anchor = south}{}
\tdplotdrawarc[color=gray,thick]{(-3.5,2.2)}{0.3}{120}{45}{anchor = south}{}
\node at (-3.49,2.35) {\large \textcolor{dark-gray}{$\cdot$}};

\draw[thick,line width = 0.6mm, color=red] (-3.5,2.2) -- (-2.93,2.77) -- (-4.07,1.63);			

\draw[thick,line width = 0.6mm, color=blue] (-5.5,1.8) -- (-5.5,1.2);					
\draw[thick,line width = 0.6mm, color=blue] (-5.5,1.8) -- (-6,1.2);
\draw[thick,line width = 0.6mm, color=blue] (-5.5,1.8) -- (-5,1.2);

\node at (-5,1.8) {\large $\theta$};
\node[color=blue, text width=3cm] at (-6.3,0.95) {LBA - receiving antenna };
\node[color=red, text width=4cm] at (-2.45,3.0) {BBOC 9217 - transmitting antenna};
\node[color=dark-gray, text width=2.5cm] at (-6.5,2.3) {Flight path};
\node at (-3.0,-0.35) {Bias Tee};
\node at (-4.5,-0.55) {\unit{24}{m}};
\node at (-1.0,2.35) {\unit{60}{m}};
\node at (-2.0,2.5) {Amplifier};
\node at (-2.0,2.2) {\unit{12}{V} DC};
\node at (-3.0,-0.7) {\unit{6}{V} DC};
\node at (-4.3,1.65) {$r$};
\end{tikzpicture}
\caption[Experimental set-up to measure the horizontal gain.]{Schematic view of the experimental set-up to measure the antenna gain pattern in a closed loop. The grey half circle denotes the path of the transmitting antenna. The angle $\theta$ indicates the zenith angle of the transmitting antenna with respect to the LBA. The black dashed-dotted line indicates the cable connection between the LBA and the amplifier used for the calibration of the network analyser.}
\label{fig: calibration setup}
\end{figure}
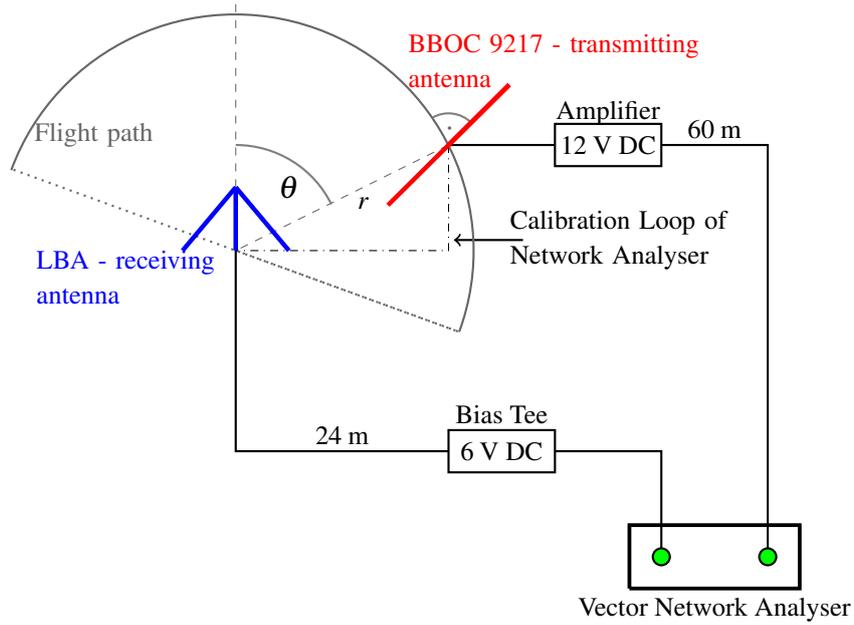

\subsection{Measurement of the antenna gain pattern}
The determination of the gain pattern of the LBA requires a study of the direction- and frequency-dependent sensitivity of the antenna to an incoming signal. Measurements between a transmitting reference antenna attached to a flying drone and a receiving LBA at different zenith angles have been performed. More details about the analysis of these measurements can be found in Chapters 4 and 5 of \cite{Krause_Thesis}.

\begin{figure}
\centering
\includegraphics[height=6cm,keepaspectratio]{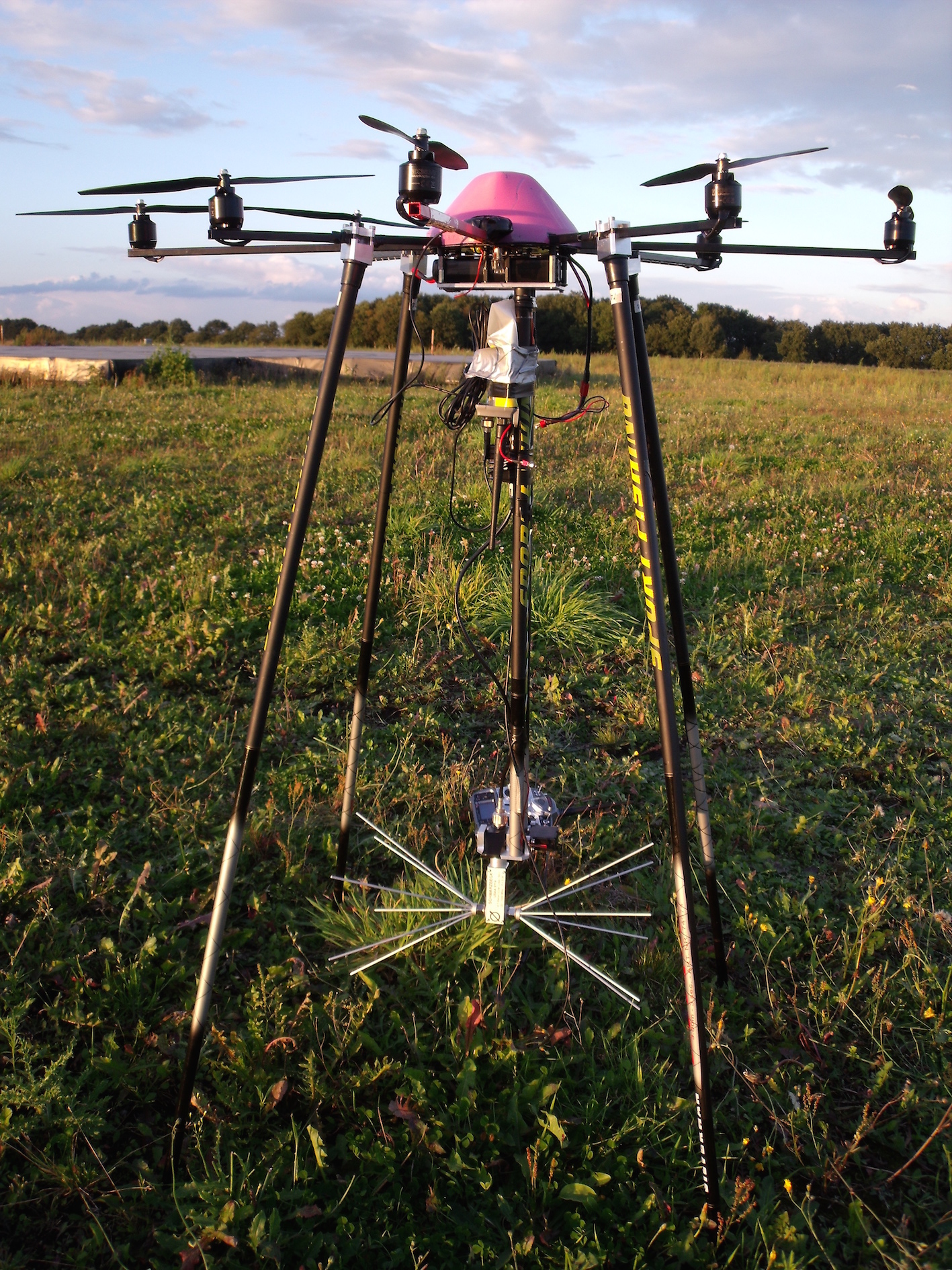}
\includegraphics[height=6cm,keepaspectratio]{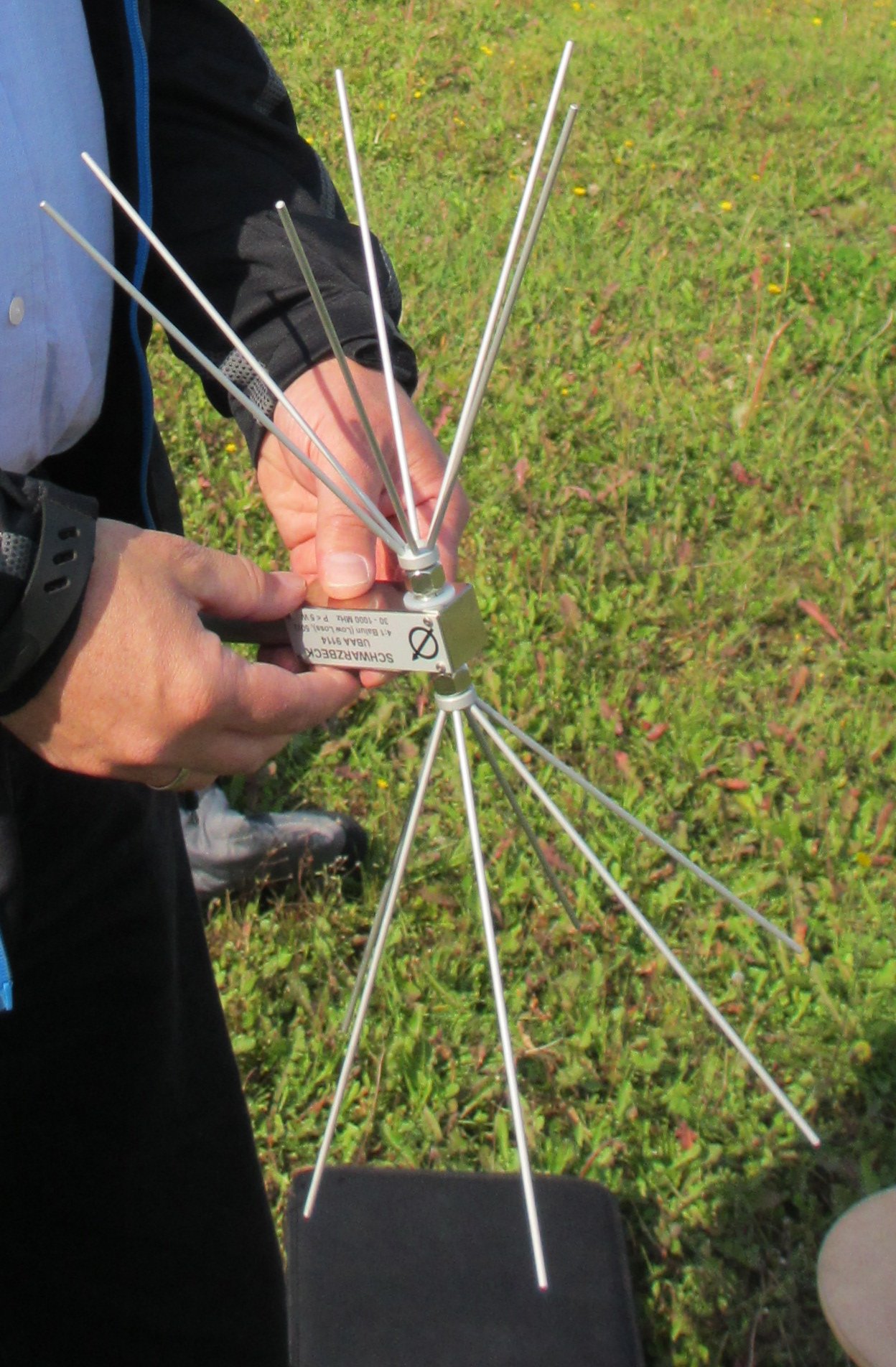}
\caption[Biconical antenna and octocopter.]{Drone and reference antenna as used at LOFAR. Left: Octocopter drone with the transmitting antenna mounted below. Right: Biconical antenna BBOC 9217 used for the calibration measurements.}
\label{Octo,antenna}
\end{figure}

\subsubsection{Experimental set-up}
The in-situ calibration measurements of the LBA have been carried out at two antennas of station CS302 in the LOFAR core during a period of two days in September 2012. The set-up requires an LBA as receiving antenna and a transmitting antenna aligned in parallel, as well as a vector network analyzer. A schematic view of the set-up to measure the zenith angle dependence of the antenna response is depicted in Figure \ref{fig: calibration setup}. 

For the measurement, a Rohde \& Schwarz FSH4 vector network analyzer with two ports was used, where one port was attached to a bias tee that supplied the receiving antenna with \unit{6}{V} DC. The other cable was directly connected to the LNA of the LBA and the received signal was measured with the vector network analyzer. Via the second port a signal was sent through the cable to the transmitting antenna attached to the drone. This signal was amplified (\unit{25}{dB}) in order to be significantly stronger than the background noise.

In order to calibrate the network analyzer, the footpoints of both antennas were connected directly. This was done only once before the actual measurements, and is represented by the dashed-dotted line in Figure \ref{fig: calibration setup}. Thereby, losses due to the electronics of the calibration set-up and additional group delay caused by the cables are directly taken into account during measurements. The vector network analyzer has a systematic uncertainty on the gain of \unit{0.6}{dB}.

For the actual measurements, the transmitting antenna has been located above the LBA with the help of an octocopter drone, with the antenna mounted below the steering electronics. 
A view of the octocopter drone with the antenna mount and transmitting antenna is shown on the left in Figure \ref{Octo,antenna}. The drone had been programmed in such a way that it flew to several pre-defined points at a certain distance to the LBA. The position of the drone was measured via a differential GPS, which enables a precise measurement of better than \unit{30}{cm} of easting, northing and height above ground every \unit{0.5}{s}. The maximum height achieved in the campaign was \unit{52}{m} above ground. This translates into a positioning uncertainty of less than $\unit{0.2}{^{\circ}}$ with respect to the arrival direction of the signal. The rotation of the drone with respect to its yaw, pitch and roll axis has been measured and corrected for, which allows us to use the same coordinate system for both antennas. A detailed description of the octocopter drone including the electronic set-up can be found in Chapter 7 of \cite{Krause2012}. 

\subsubsection{Reference antenna - BBOC 9217 Biconical antenna}
\label{sec:BBOC9217}

The BBOC 9217 antenna is an open biconical antenna developed by Schwarzbeck Mess-Elektronik \cite{Schwarzbeck}. This custom-built model uses a 4:1 balun which provides a smoother frequency response and less environmental dependence. The antenna has a mass of \unit{300}{g}, a rod length of \unit{25}{cm}, and an opening angle of $53^\circ$. 
As the flight time of the octocopter depends strongly on the overall payload, the low mass makes the antenna well suited for the calibration campaign. 
A photograph of the instrument is shown in Figure \ref{Octo,antenna}. The antenna has been calibrated by the manufacturer for a frequency range between 30 and \unit{1000}{\mega Hz}. The frequency-dependent isotropic gain of the antenna exhibits a resonance frequency of $f_{res}=\unit{(360\pm5)}{\mega Hz}$, which is outside the frequency range of the LBA. The systematic measurement uncertainty of the isotropic gain accounts to \unit{\pm1}{dB}. The reference gain curve can be found in Chapter 5 of \cite{Krause_Thesis}.

\subsection{Full signal-chain calibration}
\label{sec:crane}
The set-up as described above does not allow for an absolute calibration of the LOFAR system. The calibration loop has to be connected to the footpoint of the antenna and therefore does not include the influence of cables, bandpass filters on the RCUs and the digitization step. Therefore, an additional calibration has been performed, using a transmitting antenna that is mounted to a crane. The measurement has been performed using the same hardware as \cite{Nehls2008}, which allows us to also compare results between the different air shower experiments Tunka-Rex \cite{Kostunin:2013}, LOPES \cite{Nehls2008} and LOFAR. 
The drone calibration allows for a fast, high and very flexible positioning of the transmitting source around the receiving antenna, which provides measurements of the antenna directionality. It, however, has the draw-back that the position is not always completely stable. The calibration with a crane is more stable, but measurements of the direction dependence are less flexible and certain positions may not be reachable. 

\definecolor{dark-gray}{gray}{0.4}
\begin{figure}
 \centering
\begin{tikzpicture}[scale=1.4]
\small
\coordinate (O) at (0,0);


\draw[thick,line width = 0.6mm, color=red] (-5.5,4.) -- (-5,4.2) -- (-6,3.8);
\draw[thick,line width = 0.6mm, color=red] (-5.5,4.) -- (-5,3.8) -- (-6,4.2);
\draw[thick,line width = 0.6mm, color=red] (-5.5,4.) -- (-5,4.) -- (-6,4.);
\node[color=red] at (-3.9,4) {Reference source};
\node[color=red] at (-3.9,3.7) {VSQ 1000};

\draw[thick,line width = 0.6mm, color=blue] (-5.5,1.8) -- (-5.5,1.2);					
\draw[thick,line width = 0.6mm, color=blue] (-5.5,1.8) -- (-6,1.2);
\draw[thick,line width = 0.6mm, color=blue] (-5.5,1.8) -- (-5,1.2);
\node[color=blue] at (-2.,1.5) {LBAs + LNAs };

\draw[thick,line width = 0.6mm, color=blue] (-4.8,1.8) -- (-4.8,1.2);					
\draw[thick,line width = 0.6mm, color=blue] (-4.8,1.8) -- (-5.3,1.2);
\draw[thick,line width = 0.6mm, color=blue] (-4.8,1.8) -- (-4.3,1.2);

\draw[thick,line width = 0.6mm, color=blue] (-3.5,1.8) -- (-3.5,1.2);					
\draw[thick,line width = 0.6mm, color=blue] (-3.5,1.8) -- (-3.0,1.2);
\draw[thick,line width = 0.6mm, color=blue] (-3.5,1.8) -- (-4.0,1.2);

\draw[thick,line width = 0.6mm, color=black] (-7.2,1.2) -- (-6.7,4.8) --(-5.5,4.8) --(-5.5,4.);	
\draw[thick,line width = 0.6mm, color=black] (-7.4,1.2)--(-7.2,1.3)--(-7,1.2);
\node at (-7.5,2.9) {Crane};

\draw[dashed, line width = 0.2mm] (-5.5,1.8) -- (-5.5,3.9);
\draw[dashed, line width = 0.2mm] (-5.5,1.8) -- (-4.5,3.5);
\node at (-5.3,2.5) {$\theta$};
\node at (-5.9,2.9) {12m};

\draw[line width = 0.3mm](-5.5,1.2)--(-5.5,0.5)--(-3,0.5);
\draw[line width = 0.3mm](-4.8,1.2)--(-4.8,0.5);
\draw[line width = 0.3mm](-3.5,1.2)--(-3.5,0.5);

\draw[line width = 0.3mm] (-3,0.5)--(-3,0.8)--(-2,0.8)--(-2,0.5);
\draw[line width = 0.3mm] (-3,0.5)--(-3,0.2)--(-2,0.2)--(-2,0.5);
\node at (-2.5,0.5) {RCUs};

\draw[line width = 0.3mm](-2,0.5)--(-1.5,0.5);
\draw[line width = 0.3mm] (-1.5,0.5)--(-1.5,0.8)--(-0.5,0.8)--(-0.5,0.5);
\draw[line width = 0.3mm] (-1.5,0.5)--(-1.5,0.2)--(-0.5,0.2)--(-0.5,0.5);
\node at (-1,0.5) {TBBs};

\end{tikzpicture}
\caption[]{Schematic illustration of the experimental set-up using the reference source. The source is suspended from a crane at about \unit{12}{m} above the chosen antenna. The signal is received with the LOFAR LBA antennas and filtered and digitized at the receiver units (RCUs). The data of all antennas of a LOFAR station are read out via the LOFAR system using the transient buffer boards (TBB) as it is done for cosmic ray measurements.}
\label{fig: crane_setup}
\end{figure}
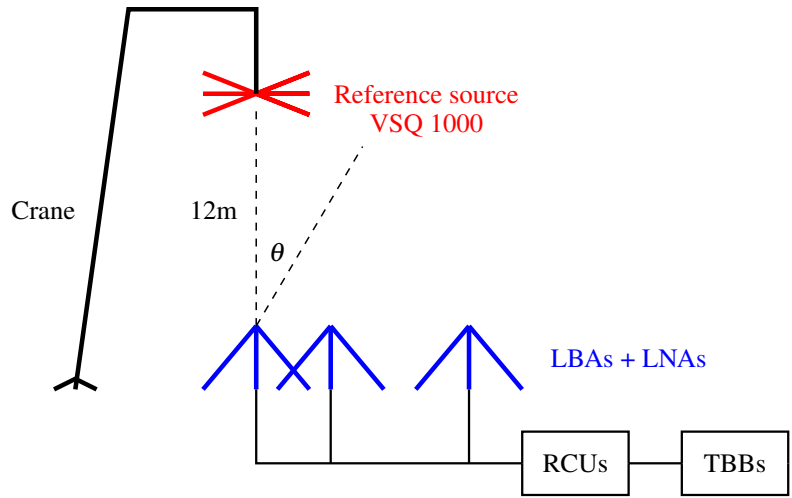

\subsubsection{Experimental set-up}
\begin{figure}
\centering
\includegraphics[height=6.5cm,keepaspectratio]{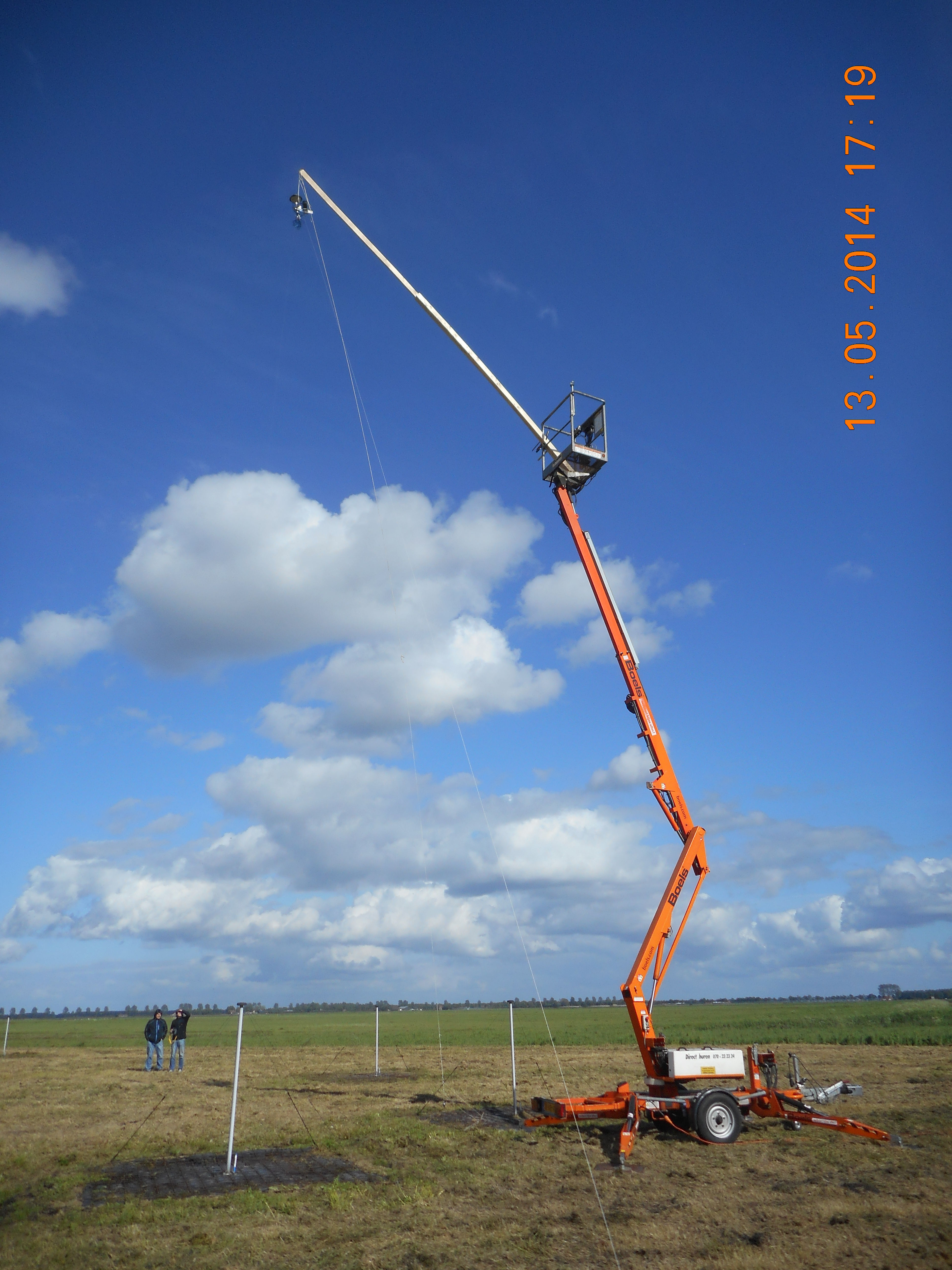}
\includegraphics[height=6.5cm,keepaspectratio]{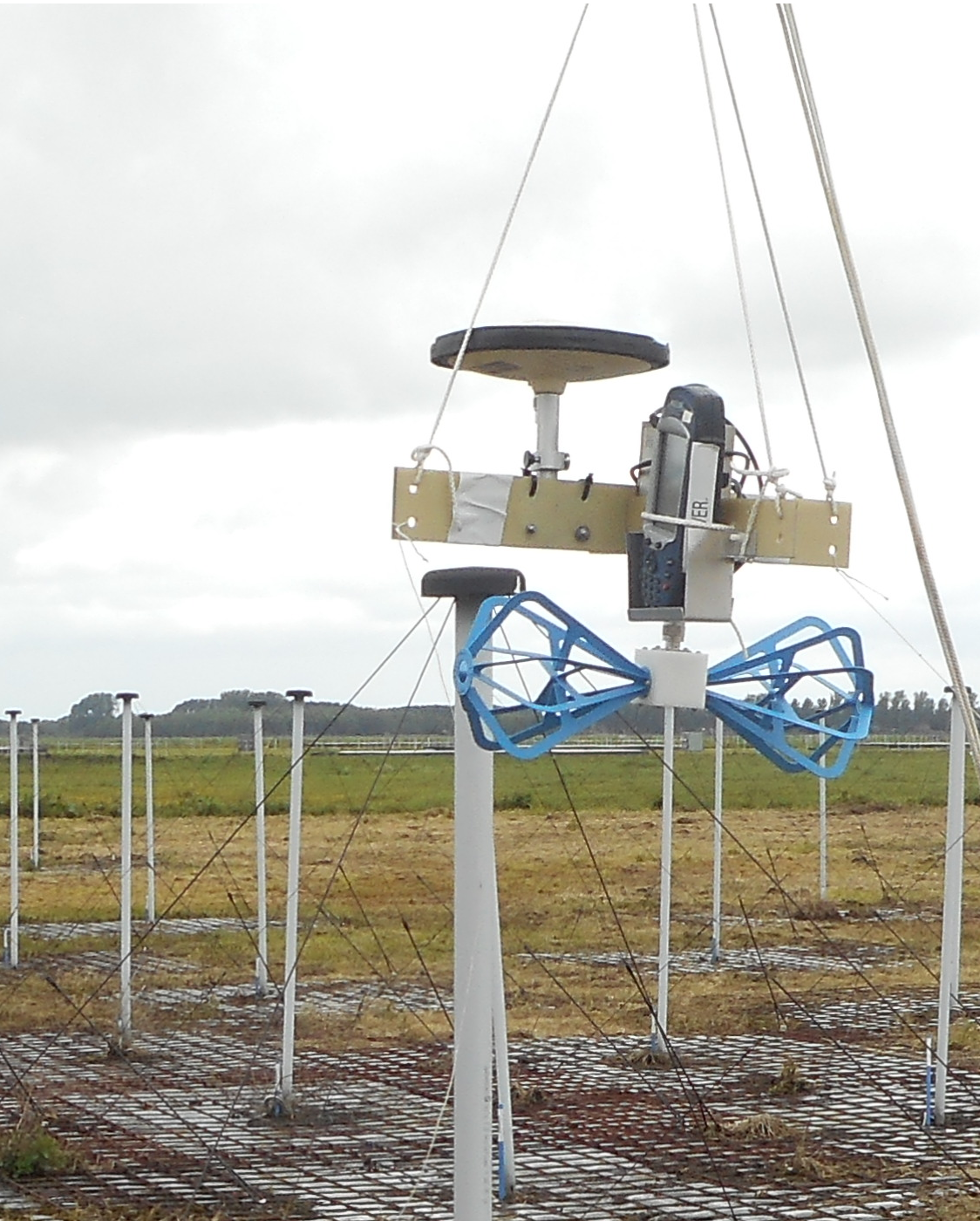}
\caption[Crane Set_up]{Experimental set-up of the measurement with the reference source attached to a crane. Left: Crane with the wooden construction and an LBA underneath. Right: Transmitting antenna as mounted on the wooden construction in front of several LBAs.}
\label{crane}
\end{figure}

The measurements with the reference source were carried out at the LOFAR superterp during a two-day campaign in May 2014. As shown in Figures \ref{fig: crane_setup} and \ref{crane}, a crane has been positioned in the antenna field of a LOFAR station close to the superterp. A wooden extension of $\unit{5.3}{meters}$ was added to the crane in order to avoid reflections from the metal construction. At the end of this extension the reference source and a differential GPS were attached. With this construction the reference antenna was positioned at a maximum distance of $r=\unit{12.65 \pm 0.25}{m}$ vertically above one dedicated antenna. The alignment of the reference antenna with a LBA dipole arm was possible with the help of two strings attached to the mount of the reference antenna as shown on the right of Figure \ref{crane}. 

For data-acquisition the LOFAR system has been used. The TBB ring-buffers of the superterp stations were read out at least 5 times per position. The final data sample, after quality cuts, consists of four read-outs, containing \unit{10}{ms} of data for each of the 48 antennas at a distance of  $r=\unit{12.65 \pm 0.25}{m}$ vertically above the central antenna. In the measured LOFAR station the same configuration is used as during air shower measurements, which ensures that the calibration includes the full signal chain. 

\subsubsection{Reference source - VSQ 1000}
\begin{figure}
\centering
\includegraphics[width=0.6\textwidth]{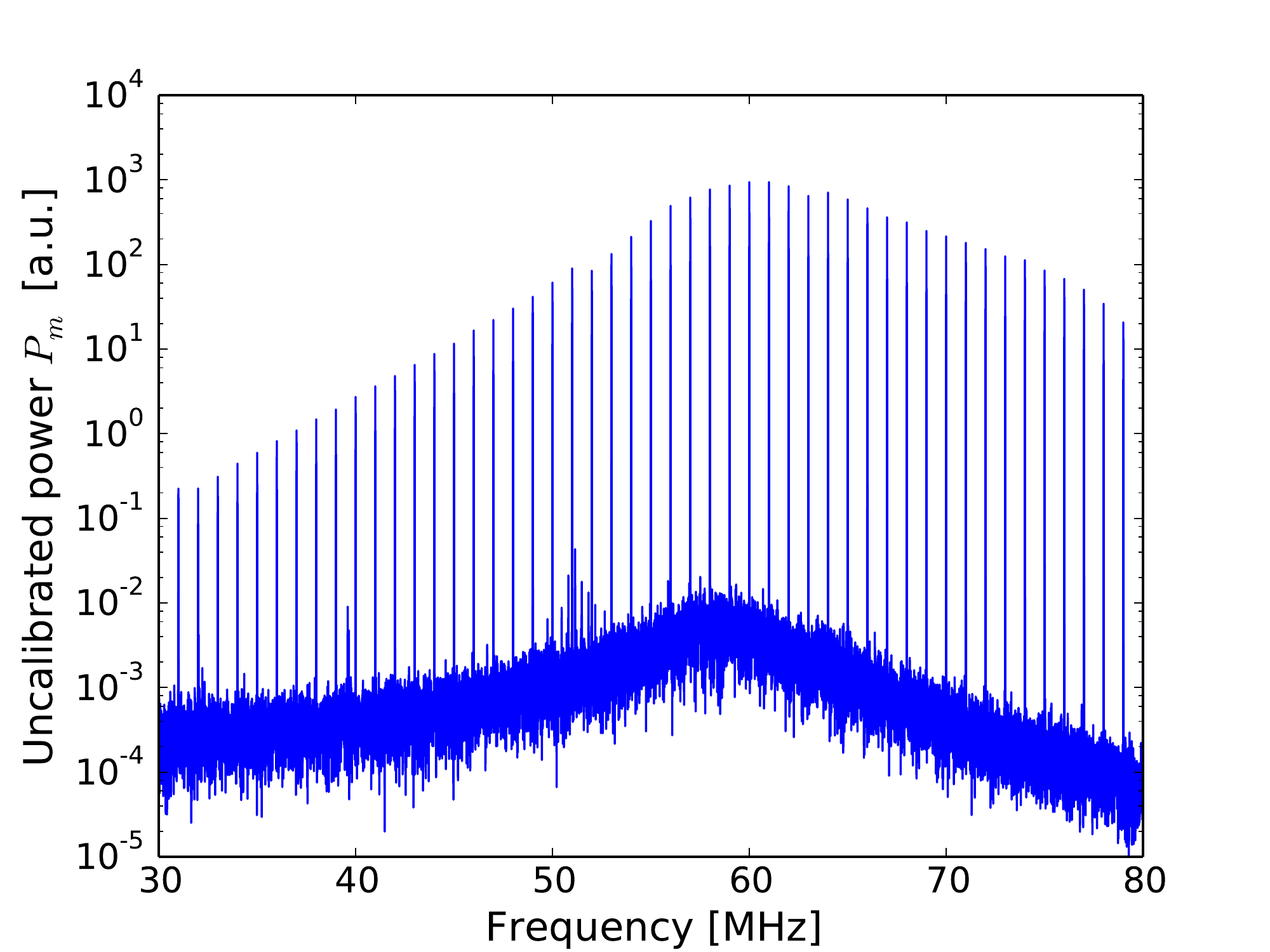}
\caption[]{Uncalibrated power spectrum as recorded with the LOFAR system for the LBA directly underneath the transmitting antenna. The power in the emitted frequencies is at least three orders of magnitude larger than the noise background of the system and the Galaxy. Single RFI lines are visible in the band of $\unit{30-80}{MHz}$. Their powers are in all cases at least a factor of 100 smaller than the power emitted by the calibration antenna and they are located at different frequencies, which makes a confusion of lines unlikely.}
\label{fig:spectrum}
\end{figure}

The reference source is a commercial product developed by Schaffner, Augsburg in Germany (now TESEQ). It is delivered as a combination of a signal generator RSG 1000 and the biconical antenna DPA 4000. The RSG 1000 is a comb-generator, generating a spectrum of single frequencies at multiples of \unit{1}{MHz} in the range of \unit{1}{MHz} to \unit{1}{GHz}. In the relevant range of \unit{30-80}{MHz} it delivers a mean power of \unit{1}{\mu W} per single frequency. It is battery-operated, which makes it ideal for measurements in the field. The DPA 4000 biconical antenna is linearly polarized and its directivity close to the main lobe follows roughly a cosine squared. This means that misalignments of less than $5^{\circ}$ with the receiving antenna result in losses of less than 1\%. The VSQ 1000 setup is certified for the 30 to \unit{1000}{MHz} frequency range in the forward direction. An example of a typical VSQ-generated spectrum detected by the LBA is depicted in Figure \ref{fig:spectrum}.

\section{Verifying the antenna model}
\label{sec:veri}
If a transmitting antenna and a receiving antenna are a loop where the impedances are perfectly matched to each other (Figure \ref{fig: calibration setup}), the gain of the transmitting antenna can be described as
\begin{equation}
G_{t}(\nu,\theta,\phi)=\frac{4\pi r^{2}S(\nu,\theta,\phi)}{P_{t}(\nu)},
 \label{eq:gain4}
\end{equation}
where $S$ is the transmitted power density  and $P_t$ the transmitted power. The gain $G(\nu,\theta,\phi)$ is proportional to the square of the VEL, $|\vec{H}(\nu,\theta,\phi)|^2$ of the antenna system. For a perfectly matched system, as it is used in the flying source calibration, both can be translated into each other \cite{Fliescher2011}. The full LOFAR antenna system, however, includes an intentional mismatch to broaden the bandwidth \cite{LBAADD},\cite{Ellingson2005}.  

The Friis Transmission Equation connects a transmitting antenna, which sends a signal, with a receiving antenna detecting this signal \cite{Friis1946} in the far-field and is used to determine the gain of an antenna. Taking into account the effective area of the antenna, the power at the receiving antenna, $P_r$ can be determined by
\begin{equation}
 P_{r}(\nu) = \left(\frac{\lambda}{{4\pi r}}\right)^{2}\ G_{r}(\nu,\theta,\phi)\ G_{t}(\nu,\theta,\phi)\ P_{t}(\nu)\ \vert\hat{a}_{r}\cdot\hat{a}_{t}\vert^{2},
 \label{eq:gain7}
\end{equation}
where $G_r$ and $G_t$ describe the gain of the receiving and transmitting antenna, respectively. 
The last factor $\vert\hat{a}_{r}\cdot\hat{a}_{t}\vert^{2}$ accounts for any possible mismatch between the polarization of the impinging wave and the polarization properties of the receiving device. If both antenna devices match in polarization and reflection, it holds that $\vert\hat{a}_{r}\cdot\hat{a}_{t}\vert^{2}=1$. If the two antennas are aligned, the gain of the transmitter $G_{t}$ is only frequency-dependent. If then $G_t$ of the transmitting antenna is known, the $G_{r}$ can be easily calculated in units of decibel
\begin{equation}
G_{r,dB}(\nu,\theta,\phi) = 20\log_{10}\left(4\pi r\right)-20\log_{10}\lambda+P_{r,dB}-P_{t,dB}-G_{t,dB}(\nu).
 \label{eq:gain10}
\end{equation}
This equation does not include influences from ground conditions or reflections on nearby antennas. Based on these equations and the flexible positioning of the octocopter, the directional and frequency dependent behavior of the antenna model can be tested. 

\begin{figure}
\centering
\includegraphics[width=0.49\textwidth]{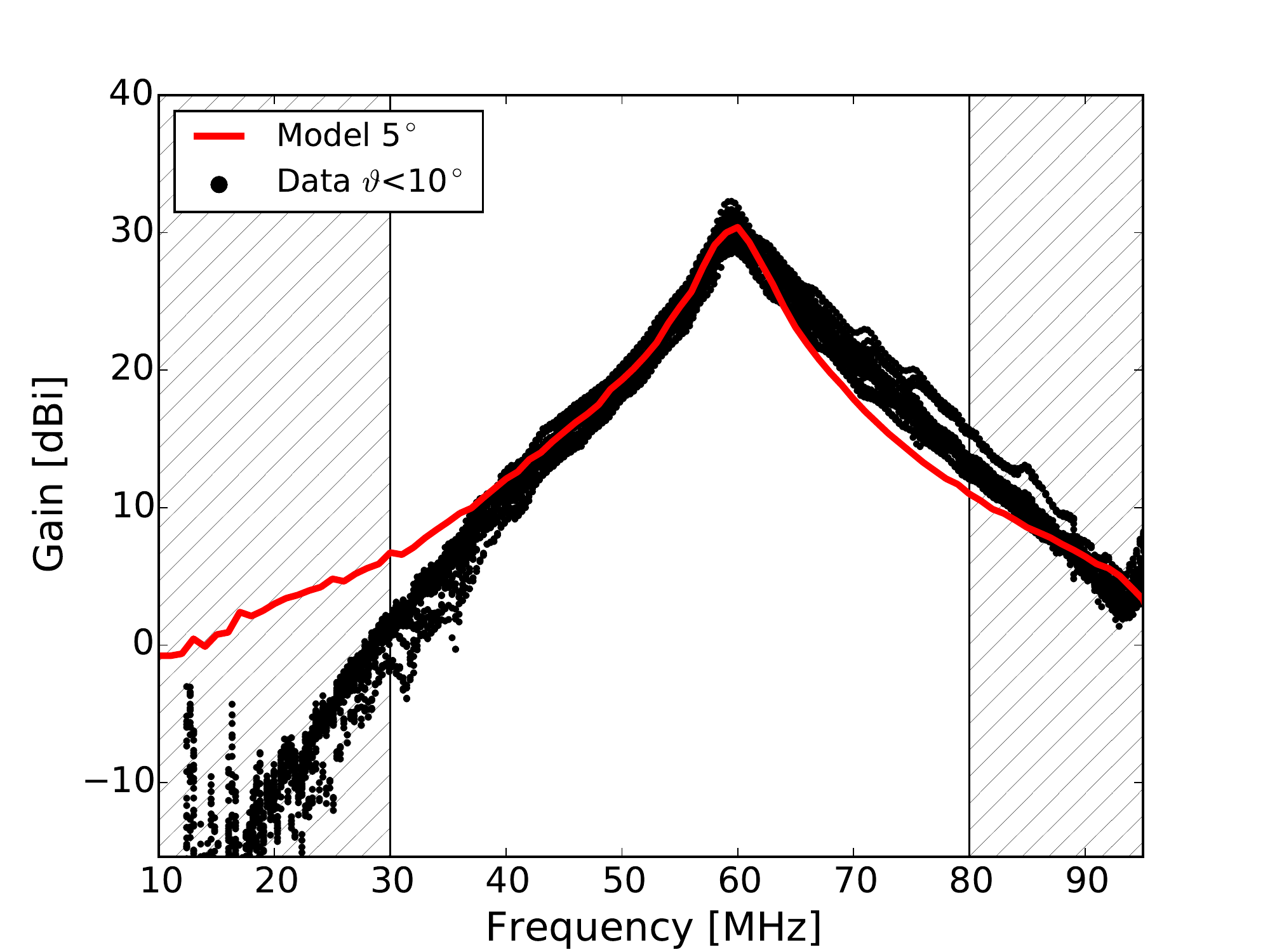}
\includegraphics[width=0.49\textwidth]{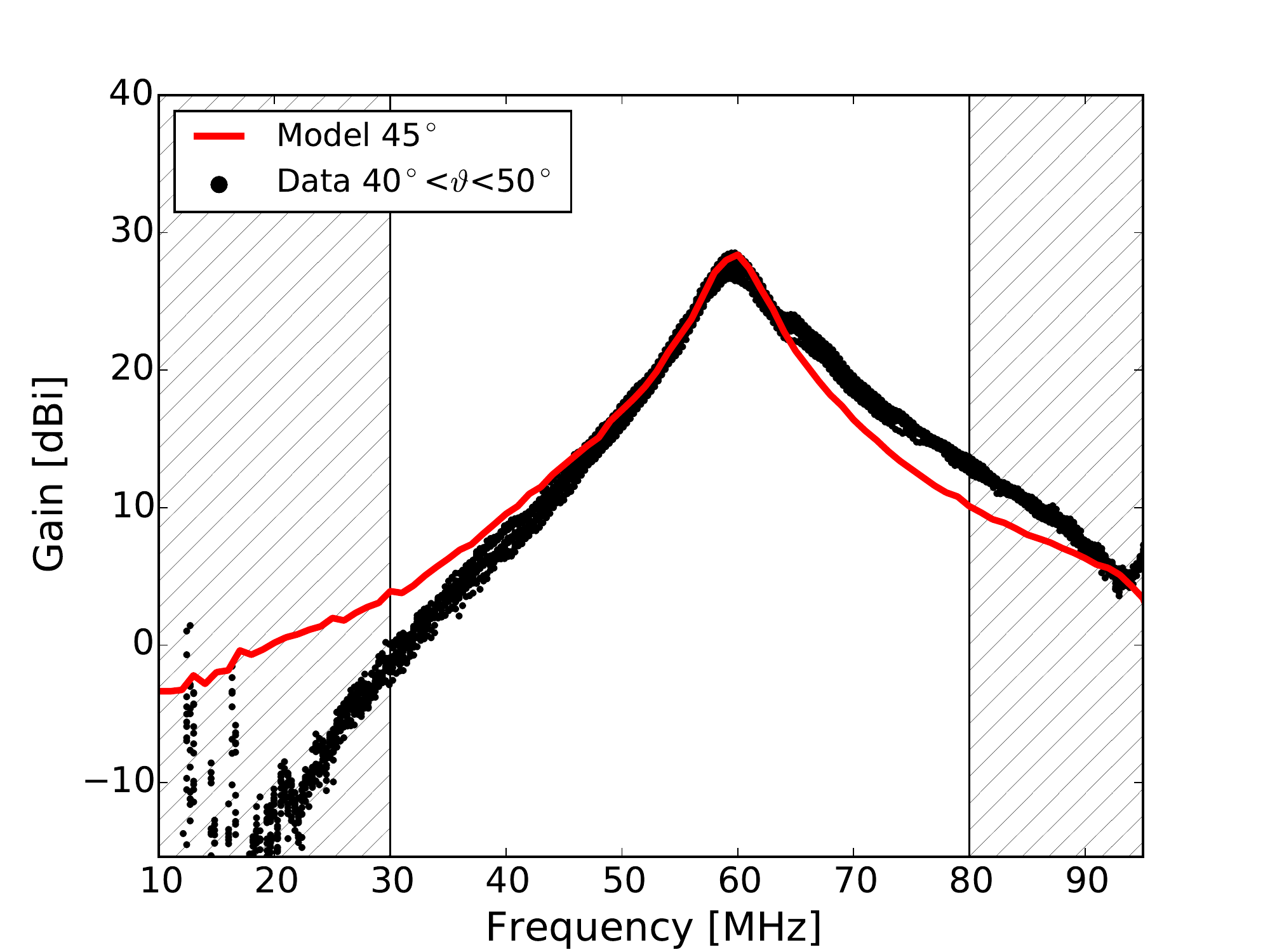}
\caption{Gain spectra as measured with an LBA in combination with the network analyzer in a loop. The spectra are shown for two zenith angle bins  for the complete LBA band of \unit{10-95}{MHz}. The left figure shows 24 spectra for $\theta< 10^{\circ}$ and the right figures contains 12 spectra in the bin of $40^{\circ}<\theta<50^{\circ}$. All measurements have been taken at the same azimuth angle. The lines indicate the gain values of the antenna model for the corresponding zenith angles of $5^{\circ}$ and $45^{\circ}$. Additional filtering outside of the band of \unit{30-80}{MHz} is not taken into account in the current model and is indicated by the hashed regions.}
\label{octo_spectrum}
\end{figure}

\subsection{Frequency behavior}
Figure \ref{octo_spectrum} shows a number of frequency dependent gain measurements for two different zenith angle bins together with the average expectations given by the antenna model. The strongest signals are measured at the resonance of the system near \unit{61}{MHz}. The figure also shows that the shape of the antenna sensitivity remains similar for different zenith bins, while decreasing in gain of about 5 dBi. 

Comparing the antenna model in detail with the measurements, it has to be noted that the simulated resonance frequency does not fully match the measured position. Foremost, outside of the central band of \unit{30-80}{MHz} the quality of the match is significantly worse. This is partly expected due to the missing modeling of the filters on the RCU. As those filters are also selectable and the frequency region is currently not used for air shower measurements, these mismatches are not investigated in detail for this analysis. Inside the central region the differences observed in the shape of the distributions are less severe, however, significant. Both issues give an indication that the simplified model of the antenna and the electronics chain that is used for the modeling for the antenna system might not be fully sufficient. A renewed simulation of the LOFAR antennas is foreseen in the future. This will involve a lot more complexity in modeling the individual components and requires a significant effort. For now and given the complexity involved, the current antenna model is kept and second order corrections to this model will be derived as part of this analysis. 

Furthermore, it should be noted that the calibration measurement is also only a snapshot of one day and time-dependent changes are hard to observe. In the complete cosmic-ray data-set, it can be observed that the resonance frequency shows small shifts in the order of up to \unit{1}{MHz}. It has been argued that the loop in the terminating wires of the LBA, as well as the ground plate accumulate water droplets. This can act as an additional impedance and change the antenna behavior. It has, for example, been observed directly that the resonance frequency shifts about \unit{0.5}{MHz} to lower frequencies after particularly heavy rain. These effects will always have to be considered as additional uncertainties.

\subsection{Directional behavior}
\begin{figure}
\centering
\includegraphics[width=0.6\textwidth]{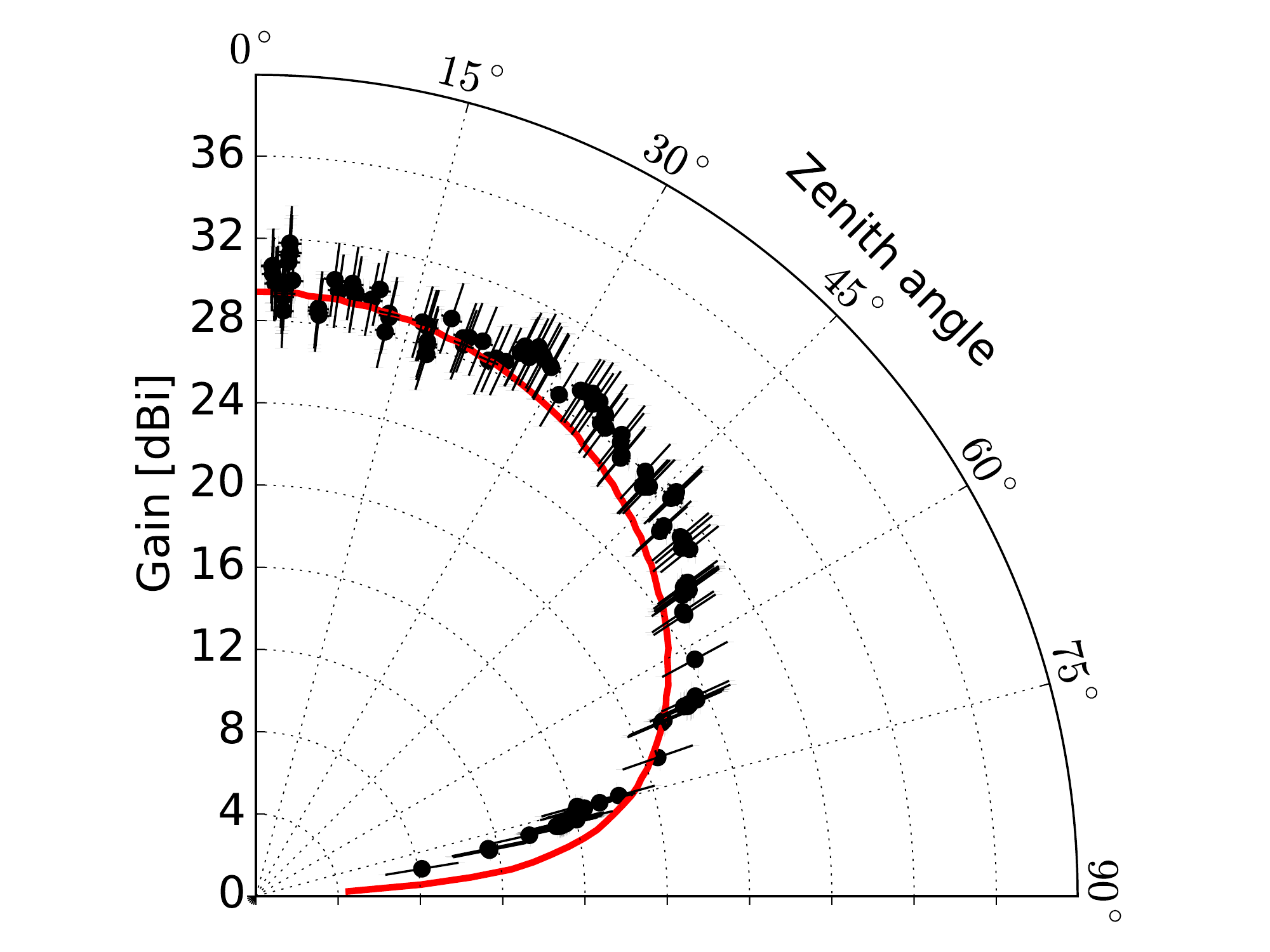}
\caption{Gain factor of the antenna as a function of the zenith angle of the incoming signal orthogonal to the antenna plane at \unit{60}{MHz}. The black points show the measurements determined from two calibration flights. The red line indicates the predicted gain of the antenna model at \unit{60}{MHz}. }
\label{octo_polar}
\end{figure}

The geometrical sensitivity of the antenna has also been tested with the same set-up. The transmission antenna has been aligned with one antenna arm at different zenith angles, which results for the emission at \unit{60}{MHz} in the pattern shown in Figure \ref{octo_polar}. The red line is the prediction of the antenna model with average ground conditions. Uncertainties on the data stem mostly from the misalignment of the transmitting antenna and the LBA and the unstable positioning. 

The simulated antenna model describes the directional behavior adequately. This is also true for the other frequency bins.
In order to estimate the influence of different ground conditions, simulations with various values of permittivity $\epsilon_{r}$ and conductivity $\sigma$ were performed. These demonstrated only a small influence ($<5$\%) on the measured gain per frequency. However, noticeable changes ($5-10$\%) on the shape of the directional dependence of the antenna pattern were observed. 

Influences due to humidity or wind are much more difficult to estimate. Since the measurement campaign only delivers a snapshot of the conditions as data were gathered at only one day with a single antenna, the campaign would need to be repeated to measure the influences, or the uncertainties have to be estimated otherwise. 

\subsection{Intermediate conclusions}
From the measurement of the antenna gain pattern, it can be concluded that the frequency behavior of the antenna model does not fully match the observed behavior. The fall-off from the resonance is different in shape in the model than what was measured during the campaign. The directional behavior is on average well described. Reasons could be either incomplete modeling of the electronics components or external influences. It should be noted that varying ground conditions or humidity do influence the modeled antenna response, especially with respect to the angular dependence. However, no combination of parameters could be found that showed an improved fit to all data, which leads us to conclude that incomplete modeling of the system is the more likely cause of the discrepancies. 

The calibration using the closed loop is not suitable for an absolute calibration, as it does not take the complete system into account. The obtained results, however, give confidence to use the current antenna model as basis and to derive additional frequency corrections based on the absolute calibration, which is described in the following sections. These corrections will have to account for possible frequency dependent damping of the cables and in the amplifiers and bandpass filters on the RCUs, as well as imperfections in the antenna model itself.


\section{Absolute Calibration of the system}
\label{sec:Abs}
For the absolute calibration two different approaches are used: one involving the calibrated reference antenna attached to a crane and the other a calibration on the diffuse Galactic radio emission, which is the dominant noise source. Both approaches are complete end-to-end calibrations, thereby including the entire receiver system, as well as cables and electronics.

\subsection{Results of the full signal-chain calibration using a reference source}
From the reference antenna (see section \ref{sec:crane}) a calibration curve can be obtained that converts the voltage traces measured with the LOFAR system  to physical units. 

  \begin{figure}
  \centering
  \includegraphics[width=0.6\textwidth]{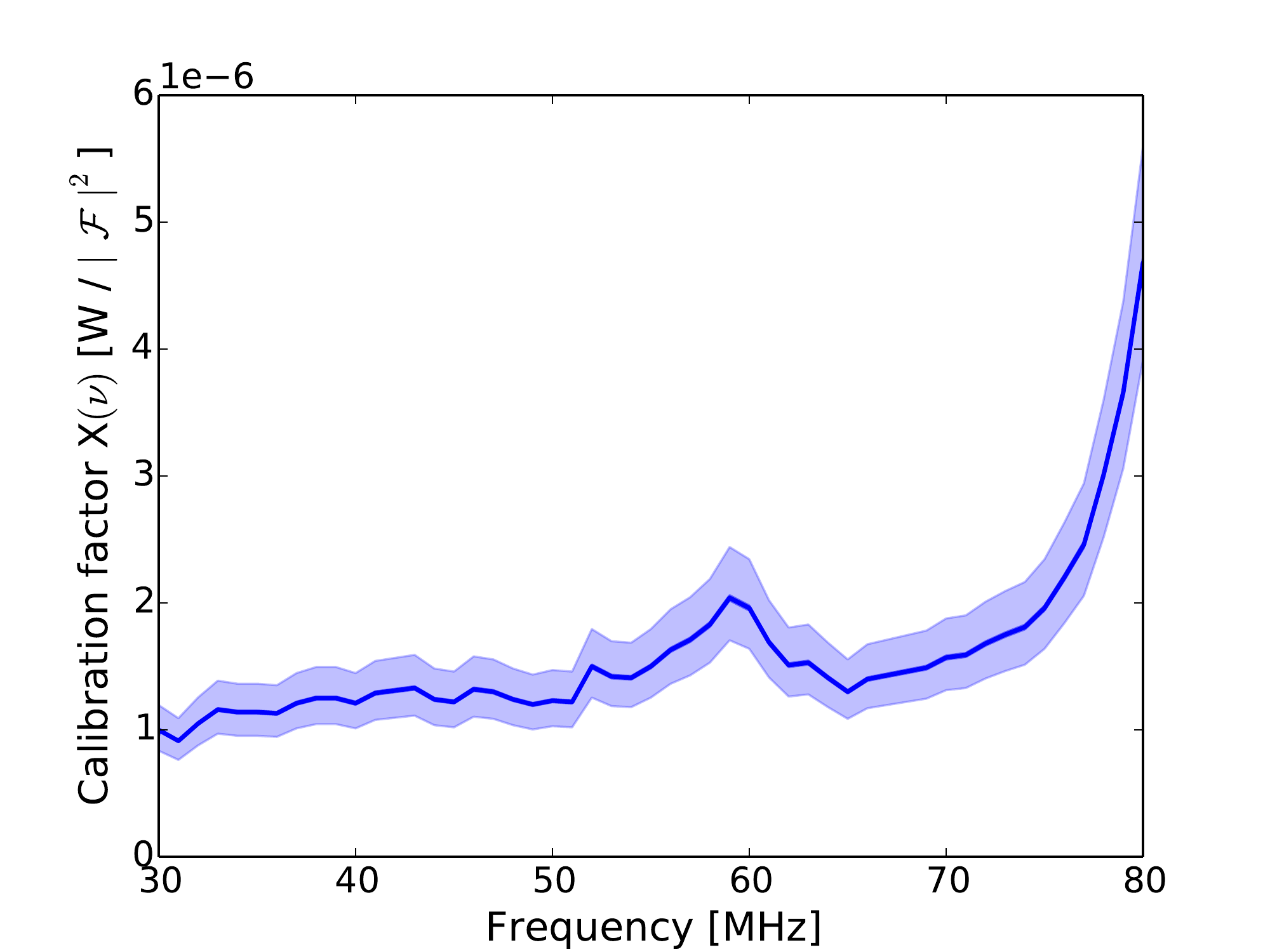}
  \caption{Calibration factor $X$ for the amplitude as a function of frequency across the LOFAR band as derived from the reference source calibration. The dark region denotes the statistical uncertainties of the method, while the lighter region illustrates the systematic uncertainties on the absolute scale.}
  \label{Calibration_curve_crane}
  \end{figure}
  
\subsubsection{Calibration curve}
The frequency-dependent calibration factor $X(\nu)$ signifies the translation between the expected power $P_e(\nu)$ in physical units and the measured power $P_m(\nu)$ in system units. The measured power in each frequency-bin is obtained from the Fourier transform of the measured analog-digital converter (AD) units as $|\mathcal{F}(\nu)|^2$. The expected power is calculated as the square of the expected voltage, divided by the vacuum impedance $Z_0$ and combined with the antenna VEL $\vec{H}(\nu)$, so that for $X(\nu)$ 
\begin{eqnarray}
X(\nu)^2 &\equiv& \frac{P_e(\nu)}{P_m(\nu)}  \label{eq:X}\\
&=& \frac{1}{Z_0} \frac{|V(\nu)|^2}{|\mathcal{F}(\nu)|^2} = \frac{1}{Z_0} \frac{|\vec{E}(\nu) \cdot \vec{H}(\nu)|^2}{|\mathcal{F}(\nu)|^2}.
\end{eqnarray}
In the case of the reference source calibration, $\vec{E}(\nu, \theta=0, \phi=0)$ is the electromagnetic field emitted by the source, the amplitude of which is obtained from the manufacturer \cite{VSQmanufacturer}. $\vec{H}(\nu)$ corresponds to the VEL from this same direction. Since the source antenna is linearly polarized, only the $J_{X\theta}$ component of $\vec{H}(\nu)$ contributes. Note that $X(\nu)^2$ is proportional to power, so that $X(\nu)$ is only proportional to the amplitude. 

For the analysis, data are used with block sizes of 65400 samples of \unit{5}{ns}, corresponding to a frequency resolution of~$\sim$\unit{3}{kHz} in the \unit{1-100}{MHz} range. The bandwidth is restricted to \unit{30-80}{MHz}. A Gaussian smearing of the edges affecting 5 frequency bins at both ends of the spectrum has been applied to the filter to reduce sharp cut-off effects. Signal peaks from the comb generator have a width of less than 9 kHz, corresponding to at most 3 frequency bins with this resolution. The background noise, as well as single narrowband noise-lines are at least three orders of magnitude lower than the signal, and therefore contribute less than 1\% in power. 

The resulting calibration factor $X(\nu)$ is shown in Figure~\ref{Calibration_curve_crane}. The curve is relatively constant over most of the frequency range, with a clear steepening at the upper range of the spectrum. This suggests that the antenna response is flatter at the highest frequencies than what is currently modeled. This confirms the finding of the antenna gain measurement, as shown in Figure \ref{octo_spectrum}. Additionally, a clear peak is visible around  \unit{59}{MHz}. This is likely caused by the already discussed shift in resonance frequency between data and antenna model. The measured peak shifts in correlation with environmental changes, such as rain, and it is therefore rather challenging to match this correctly in a model.  Different measurements conducted on the same day show stable results with variations of about $\sim$1\%. Uncertainties are however dominated by systematic uncertainties in the emitted power of the VSQ.

  \begin{figure}
  \centering
  \includegraphics[width=0.6\textwidth]{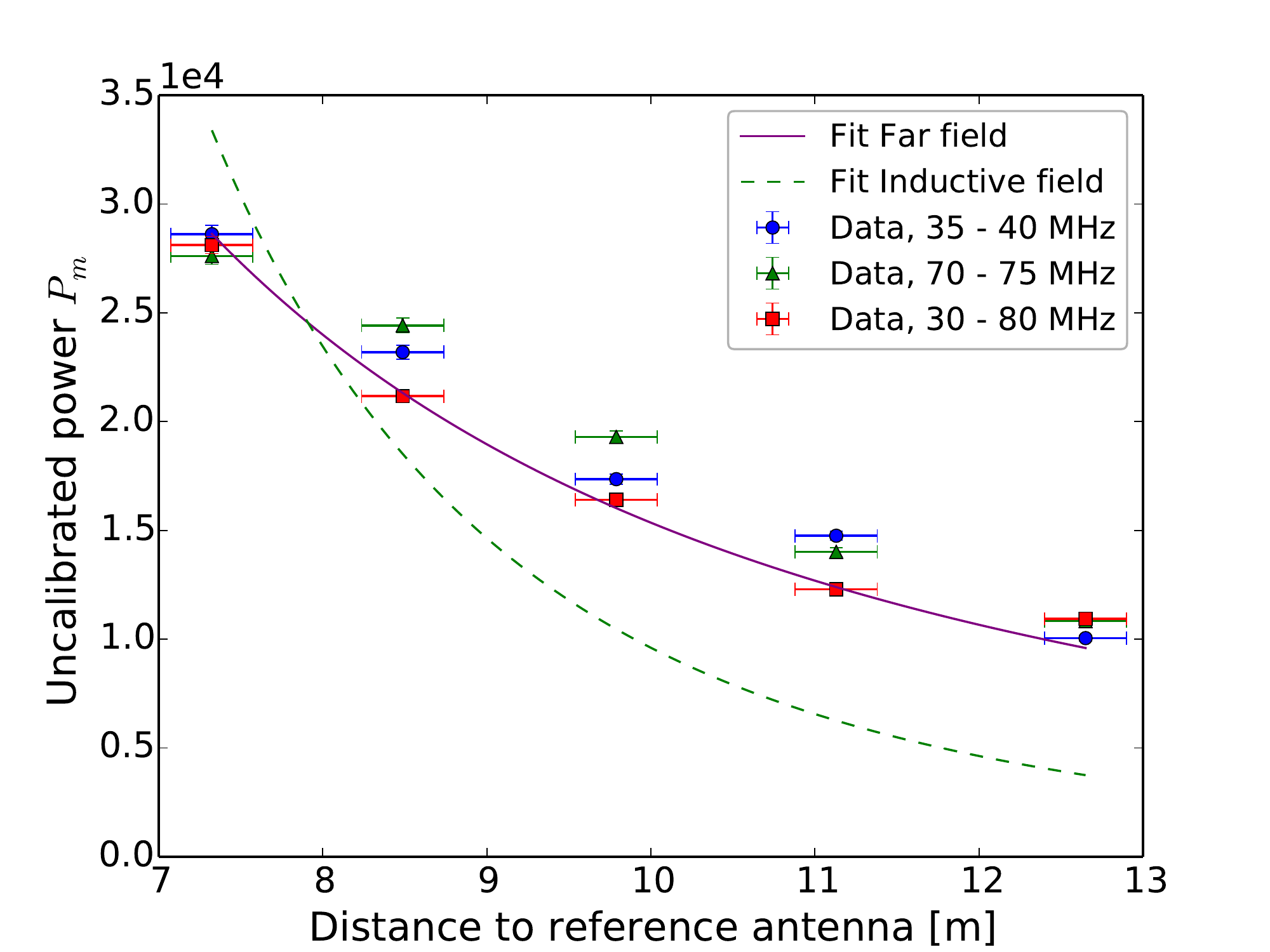}
  \caption{Integrated uncalibrated power received in different frequency ranges (markers) as a function of distance between reference and receiving antenna. Fitted to the data (full bandwidth) is the power fall-off for either a dominating inductive field (dashed line), or for a dominating far field (solid line).} 
  \label{Farfield}
  \end{figure}

\subsubsection{Uncertainties and cross-checks on calibration}
Several uncertainties have to be considered and are given, in the following, as relative uncertainty $x$ on $X(\nu)$, meaning that the $1\sigma$ confidence-interval is described by $[X(\nu)/(1+x),X(\nu)\times(1+x)]$.

The largest contribution to the uncertainties follows from the source antenna, which is dominated by a systematic uncertainty of 16\% on the emitted electric field strength as given by the manufacturer. Signal-instabilities caused by fluctuations in temperature introduce another systematic 6\% on the emitted electric field strength \cite{Nehls2008}. This contribution is expected not to be present during the measurement due to the same type of weather throughout the campaign, but will affect the applicability of the calibration at other environmental conditions. 
Additionally, possible misalignments between the source and the receiving antenna influence the result. Offsets are estimated to be less than \unit{0.25}{m} in altitude and 5$^\circ$ in rotation, resulting in uncertainties on the calibration factor $X(\nu)$ of up to 5\%. Considering all influences of the measurement setup, such as antenna cross-talk and signal reflections on nearby surroundings, is believed to yield uncertainties of no more than 5\%. 

For an accurate approximation of the signal, the source needs to be located in the far field where the power $P$ falls off as $P \propto r^{-2}$ as discussed in section \ref{sec:set-ups}. The determining distance for our case is $r >> \lambda$, where $\lambda = 10$ m for \unit{30}{MHz}. This questions whether our measurements are truly in the far-field.  However, the integrated measured powers reveal a good fit to a far field approximation for all frequencies even at distances $r < \lambda$  as shown in Figure \ref{Farfield}. While this suggests the correct use of a far field approach, spill-over effects from the near field cannot be excluded completely.  Therefore, an additional uncertainty originating from a far-field approximation is approximated to be 0.5~dB~($\sim$~6\% in amplitude) \cite{AugerAntennas2012}.  
  
During the calibration measurement the additional 47 dual-polarized antennas were also read out in the LOFAR station. This could allow one to cross-check the calibration value and check for uncertainties between antennas. However, in order to be able to do so, one needs to make an assumption about the emission pattern of the calibration source since the manufacturer only provides values for the main lobe. This additional antenna model increases the complexity of the test. Differences between the expectation and the measurement can either be attributed to issues with the model of the LBA or the one of the source antenna and neither can be excluded. As a first test indeed shows an unexpected but smooth trend in the calibration curves, we have refrained from using receiving dipoles not directly under the reference antenna. 

\begin{table}[t]
    \begin{center}
       \begin{tabular}{l l l r }
         \hline
         \hline
          Uncertainty ($\sigma$) &&& Value [\%] \\
         \hline
         \hline
         {\bf Antenna-by-antenna}  && Variations between antennas          & 1 \\
         \cline{2-4}

         && {\bf Total}  &{\bf 1 }\\
&&& \\
         \hline
         \hline
         {\bf Event-by-event}  && Environmental (excl. source)          & 5 \\
         \cline{2-4}

         && {\bf Total}  &{\bf 5 }\\
&&& \\
         \hline
         \hline
 {\bf Calibration}                           & Method-specific &  Source emission & 16 \\
                              &                  & Far-field & 5 \\
                               \cline{2-4}

                           && {\bf Total}  &{\bf 17 }\\
&&& \\
                            &                    & Set-up & 5\\
           & Campaign-specific & Alignment           & 5 \\
                            &                   & Source temperature stability & 6\\
                            &                 & Measurement variations & 1 \\
         \cline{2-4}

         && {\bf Total}  &{\bf 9 }\\
          \hline
          \hline
       \end{tabular}
    \end{center}
    \caption{Summary of the uncertainties on the calibration curve in amplitude that have to be considered for the full signal-chain calibration using the reference antenna.}
     \label{tab:calcrane_uncertainties}
\end{table}

When comparing the integrated power between different measurements on the same day, the statistical fluctuations are less than 1\%. From the Galactic calibration (see section \ref{sec:galactic_results}), which can take into account all antennas, we have estimated that the variation between antennas is about 1\%, if the antenna is fully functioning, which is the case for the measured antenna. Additionally, also following the Galactic calibration, environmental effects on the LBAs are determined to contribute 5\% uncertainties. An overview of all different contributing uncertainties can be found in Table~\ref{tab:calcrane_uncertainties}. Using conservative estimates the total uncertainties sum up to 
\begin{eqnarray}
 \frac{\sigma_{X(\nu)}}{X(\nu)} = &\pm& 1\% \;(\text{antenna-by-antenna}) \pm 5\% \;(\text{event-by-event}) \\
 &\pm& 17\% \;(\text{method-specific}) \pm 9\% \;(\text{campaign-specific}).
\end{eqnarray}
More details about this analysis can be found in \cite{Tijs}.


\subsection{Results of the calibration using the Galactic emission}
\label{sec:galactic_results}
The LBAs have a large field of view and data collected with the system are known to be sky noise rather than system noise dominated. With the cosmic-ray data collected since 2011 a precise and continuous reference of the noise floor is available which is used for calibration purposes. 

The measured noise is a combination of sky and system noise. The system noise temperature $T_{sys} (\nu)$ is defined as the sum of the receiver noise temperature $T_{rec}(\nu)$ and the electronic noise temperature $T_{elec}(\nu)$:
\begin{equation}
T_{sys} (\nu) = T_{rec} (\nu) + T_{elec}(\nu) \label{eq:tsys}.
\end{equation}
The latter combines all noise contributions from filters, amplifiers, cables and receivers. In the absence of RFI, the receiver noise temperature only contains emission from the Galaxy, so that $T_{rec}(\nu) = T_{sky}(\nu)$~\cite{RCUmenno}.
The combined noise floor of electronics and sky, $T_{sys}$, is compared to predictions from models of the diffuse emission of the Galaxy in order to obtain an absolute calibration.

\subsubsection{Predictions of the Galactic emission}
In this analysis, two different models are used for the prediction of Galactic radio emission; LFmap \cite{LFmap} and  the Galactic Sky Model (GSM) \cite{GSM}. Both work by interpolating published reference sky-maps in the tens of MHz to several GHz range to any desired frequency in that range, and calculate a sky map $T_{sky}(\nu, \alpha, \delta)$ with $(\alpha, \delta)$ being equatorial coordinates. The two models differ by less than 5\% in the LBA bandwidth and quote similar uncertainties. Two types of uncertainties have to be distinguished. Both models fit the reference maps, which means that there is a contribution of the quality of the fit and of the absolute scale that the fits inherit from the original measurements. The fit is reported to describe the underlying maps with an accuracy of usually $5\%$ and only for some maps with up to $10\%$ across the whole fitted range of frequencies.  

The two models directly inherit the absolute scale of the reference maps and can be only as good as the data. The highest quoted uncertainty of the original data (20\% in power) \cite{GSM,1977A&A....61...99B} is therefore assumed as a conservative estimate for the overall scale uncertainty. 
This conservative estimate also accounts for the fact that the LOFAR bandwidth has been less widely covered in surveys and some reference maps are incomplete for a few regions with high emission. Future dedicated work might be able to reduce these uncertainties to less conservative values. 

\begin{figure}
\centering
\includegraphics[width=0.6\textwidth]{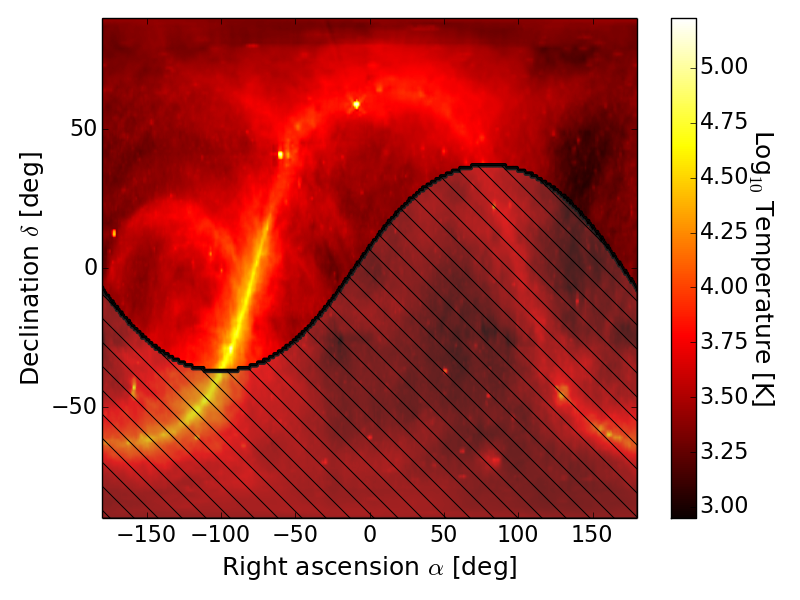}
\caption{Map of the Galactic radio emission, generated by LFmap \cite{LFmap} at 60 MHz. The upper region is visible from the location of LOFAR at 17.3 hours Local Sidereal Time (LST). The black line corresponds to the horizon and the dashed region is not visible at this specific time.}
\label{LFmap_example}
\end{figure}

As differences between LFmap and GSM are relatively small, only the LFmap model of the Galactic radiation is used for the analysis. An example of a map of the emission at 60 MHz as obtained from LFmap is shown in Figure~\ref{LFmap_example}.

The power expected from Galactic radio emission can be calculated via the Rayleigh-Jeans law. The received spectral power~$P_{\nu,\;X}$ in one of the LOFAR dipoles (X)  is given by
\begin{eqnarray}
P_{\nu,\;X} &=&  \frac{2 k_B}{c^2} \; \nu^2 \int_\Omega T(\nu, \theta, \phi) \;|\vec{H}(\nu, \theta, \phi)|^2 \;d\Omega \\
<P_{\nu,\;X}> &=& \frac{k_B}{c^2}    \; \nu^2 \int_\Omega T(\nu, \theta, \phi) \;(|J_{X\theta}|^2 + |J_{X\phi}|^2) \;d\Omega. \label{eq:px}
\end{eqnarray}
using equation \ref{eq:jones}. More precisely, equation~\ref{eq:px} follows as a result of a time-averaged antenna response, determined by the received voltage $V_X$ in a dipole X as the response to unpolarized waves
\begin{eqnarray}
<V_X^2> &=& <(\vec{E}(\nu, \theta, \phi) \cdot \vec{H_X}(\nu, \theta, \phi))^2> \\
&=& <(J_{X\theta} \; \sin(\omega))^2 + (J_{X\phi} \; \cos(\omega))^2 + 2 \cdot J_{X\theta} J_{X\phi} \; \sin(\omega) \; \cos(\omega) > \\
&=& \frac{1}{2}(|J_{X\theta}|^2 + |J_{X\phi}|^2). \label{eq:Jones}
\end{eqnarray}
The antenna-based coordinates $(\theta, \phi)$ are calculated from $(\alpha, \delta)$ using the Local Sidereal Time, and the location of the LOFAR superterp at latitude \unit{52.92}{^\circ N} and longitude \unit{6.87}{^\circ E}. As the visible sky region changes over time, the power received from the Galaxy is also time-dependent. If such a variation is observed in the data, the system is sensitive to and likely dominated by the sky noise. 

         \begin{figure}
          \centering
          \includegraphics[width=0.495\textwidth]{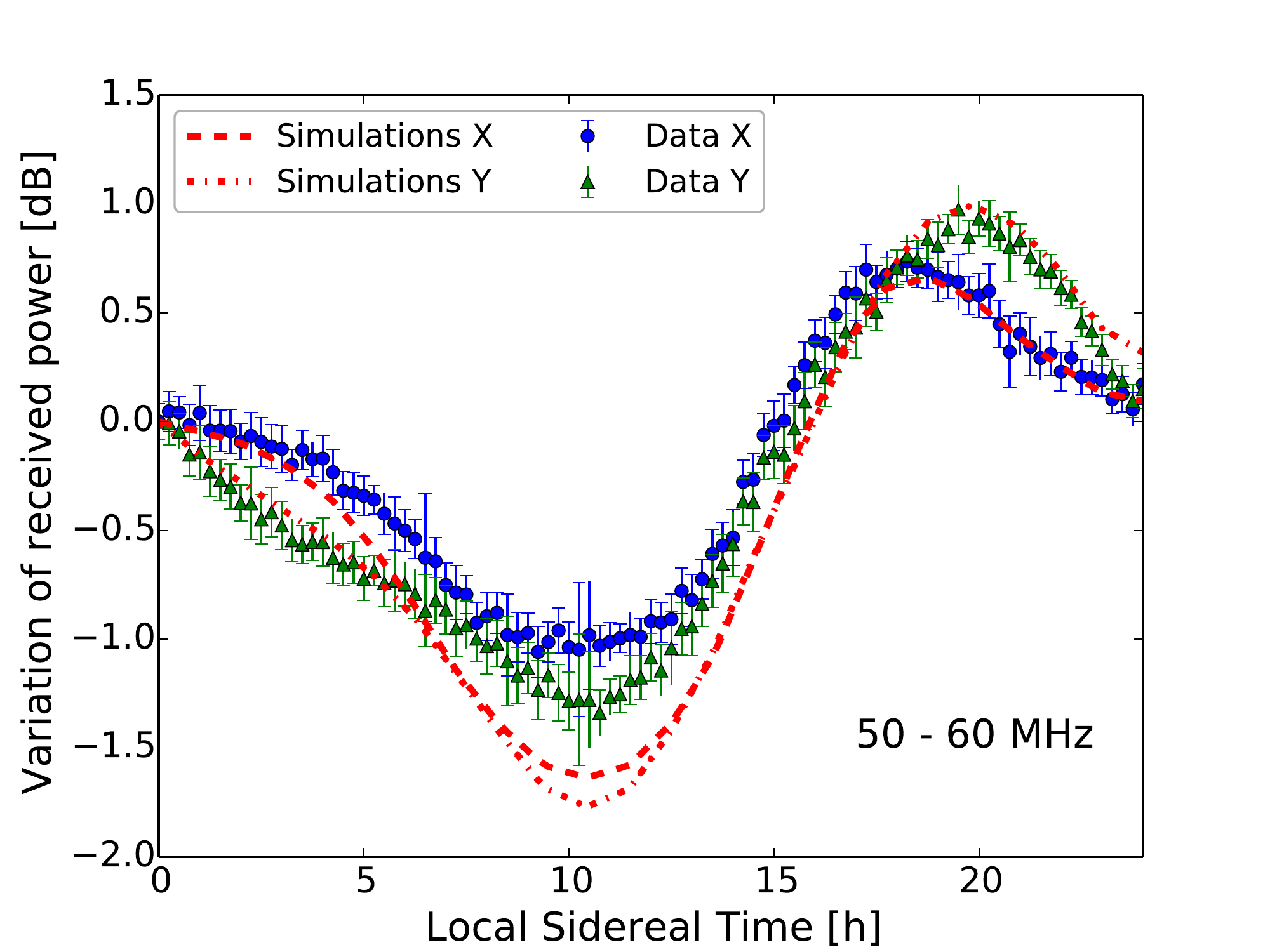}
          \includegraphics[width=0.495\textwidth]{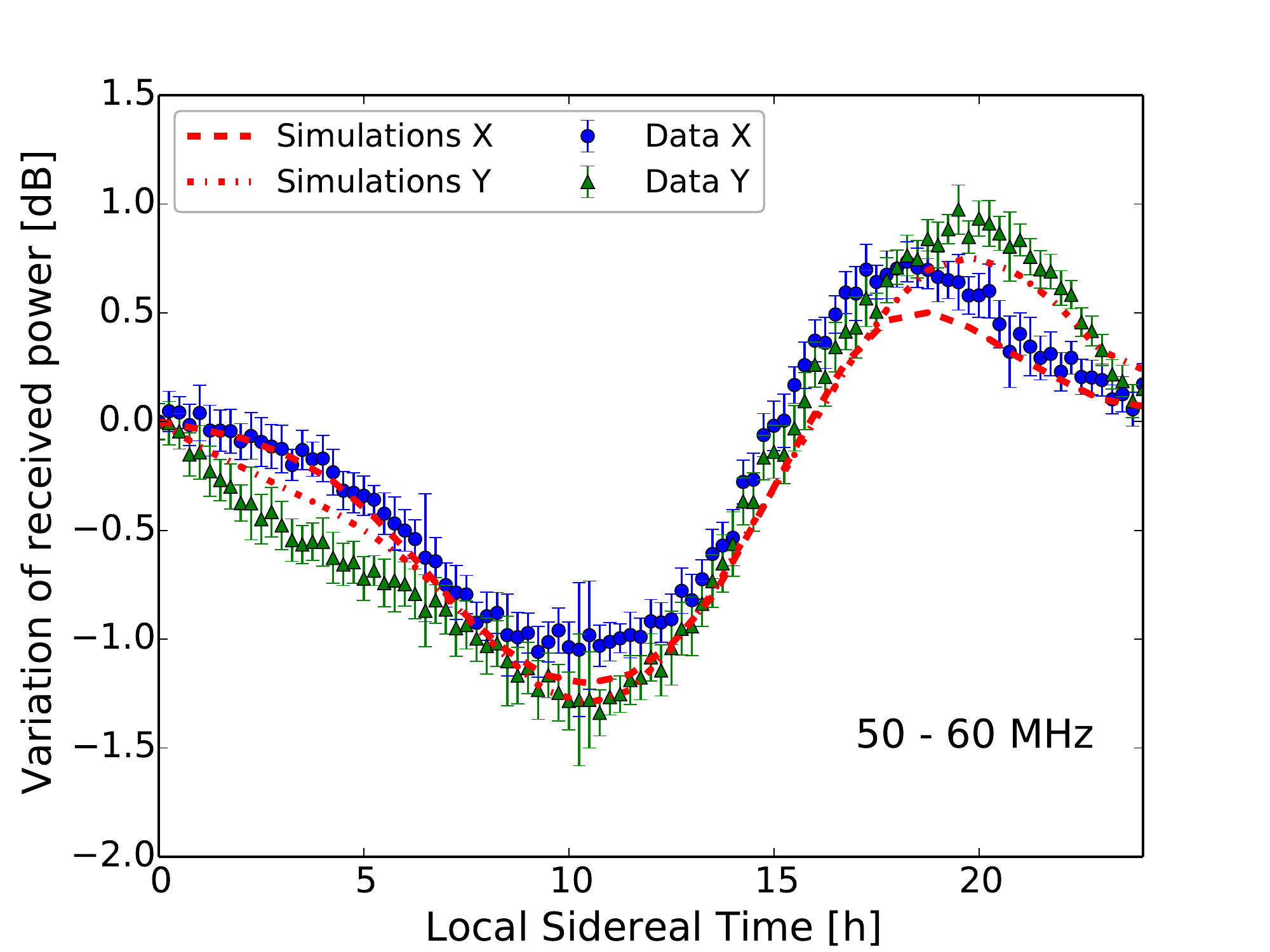}
          \caption{Integrated median uncalibrated noise power, as a function of the Local Sidereal Time, for the $\unit{50-60}{MHz}$ sub-band. Also shown is the predicted received power in both dipoles (dashed red lines), before (left) and after (right) applying electronic noise corrections.}
          \label{fig:galaxy_sim_and_data_compared_corrected}
          \end{figure}

\subsubsection{Contribution of the electronic noise}
The measured power is depicted in Figure~\ref{fig:galaxy_sim_and_data_compared_corrected} as a function of Local Sidereal Time. A clear variation is visible. In order to match simulated powers from the Galactic emission to the detected data, the contribution from the electronic noise needs to be included. The system noise has been measured during deployment and commissioning \cite{MennoRCUmeasurement,ScaifeHeald2012,vanHaarlem2013}. All those measurements, however, only intended to show that the system is sky noise dominated and did not focus on providing an uncertainty. Thus, our data are used to directly determine the most probable fraction of sky noise. 

The most probable noise offset due to the electronic noise has been found using a least-squares fit. The data was binned in LST-bins of about 15 minutes. With the current complete data-set every bin then contains the noise background of an average 34 air showers per antenna. As the electronic noise is frequency-dependent, offsets need to be determined in frequency sub-bands. Here, sub-bands of $\unit{10}{MHz}$ (i.e. ranging [$30 - 39, 40 - 49, \ldots, 70 - 79$] MHz) were chosen. The reduced $\chi^2$ was calculated for every combination of electronic noise offset per frequency band and the simulations of the Galaxy with respect to the binned data. The uncertainties per sub-band correspond to noise corrections at the point where $\chi^2 = \chi_{min}^2 + 1$, which is a conservative estimate of the uncertainty. 
        \begin{figure}
        \centering
        \includegraphics[width=0.6\textwidth]{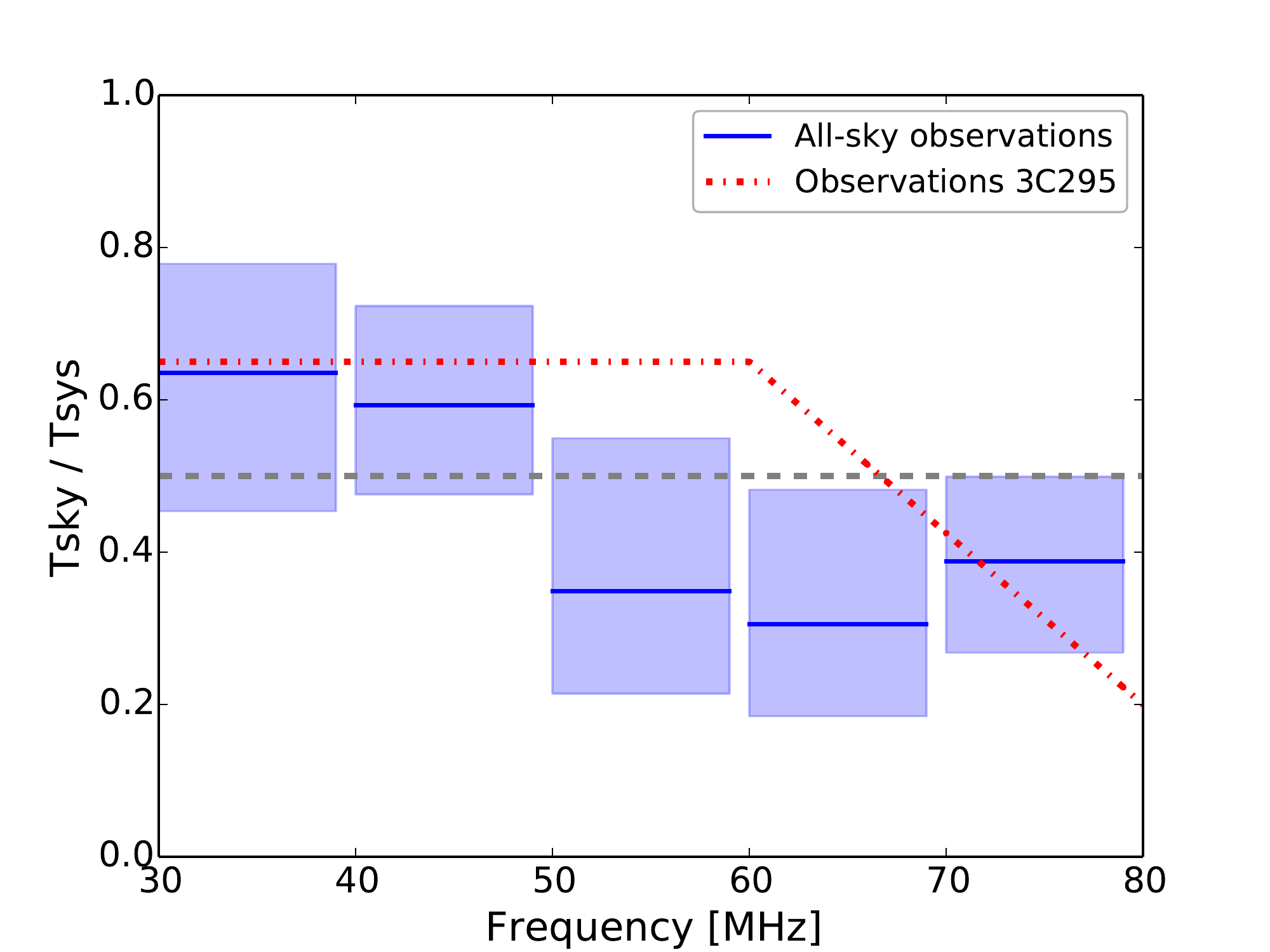}
        \caption{Calculated values for the contribution of the sky temperature to the total system temperature. Shown are both the results from this analysis  (blue lines with shaded uncertainty) and an approximation of the median values derived from the observation of 3C295 \cite{vanHaarlem2013} (dashed dotted red line). The dashed line indicates a fraction of 0.5. }
        \label{Tsky_Tsys}
        \end{figure}

With these values, a measure for $T_{sky} / T_{sys}$ can be constructed per sub-band $\nu_0$ as 
\begin{equation}
\resizebox{.3\textwidth}{!}{$\displaystyle 
\frac{T_{\nu_0, \; sky} }{T_{\nu_0, \; sys}} = \frac{T_{\nu_0, \;sky}}{  (\frac{V_{\nu_0, \;noise}^2}{J_{30 \; - \; 80, \;mean}^2} )  +  T_{\nu_0, \;sky}       }. 
$}\label{eq:TsysTsky}
\end{equation}
\noindent
Here, the temperature $T_{\nu_0, \; sky}$ in a certain sub-band $\nu_0$ is determined as the average sub-band voltage divided by the average VEL amplitude of the antenna model in each sub-band. The electronic noise offset per band is divided by the average value of the VEL in the full $\unit{30 - 80}{MHz}$ range for normalization. Resulting values are depicted in Figure~\ref{Tsky_Tsys}. Values are comparable to what has been established earlier \cite{vanHaarlem2013}. The largest discrepancy is near the resonance frequency, which is most easily affected by using a slightly different antenna model. As the uncertainties on the astronomical method can no longer be obtained, no significance of this discrepancy can be given.

The variation of both measured spectral power and predicted power before and after noise corrections are shown in Figure~\ref{fig:galaxy_sim_and_data_compared_corrected}, for the \unit{50 - 60}{MHz} sub-band, where LOFAR is most sensitive due to the resonance frequency of the dipole.  The figure shows that the predicted curves match the data better after the noise correction. Subtle systematic differences indicate that an improved understanding of the contribution of the  electronic noise and the antenna model can positively affect this analysis. A similar behavior can be observed for the other frequency bands. The obtained noise corrections have been added to the simulations of the sky noise for all further analyses. It is foreseen to measure the noise contribution in a dedicated measurement campaign to reduce the uncertainties. 

\subsubsection{Calibration curve from comparison to data}
In a way identical to the reference source calibration (eq.~\ref{eq:X}), a calibration curve for the full signal-chain can be established for the predicted Galactic emission using the same antenna model. The expected power is composed of both Galactic radiation and electronic noise combined, while the measured power in each frequency bin is again calculated as $|\mathcal{F}(\nu)|^2$. 

\begin{figure}
\centering
\includegraphics[width=0.6\textwidth]{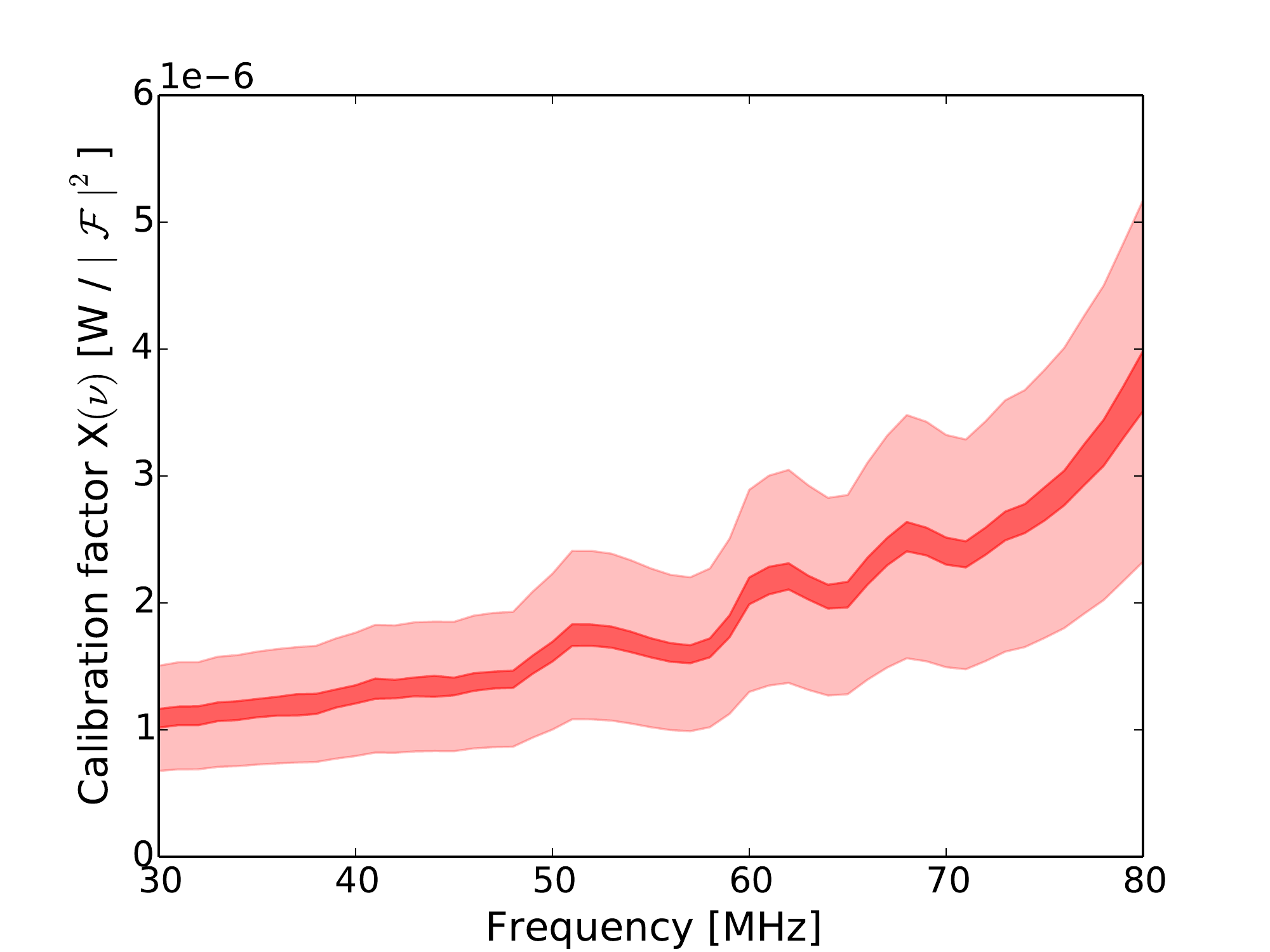}
\caption{Calibration factor for the measured amplitude as a function of the frequency as derived from diffuse Galactic emission. The dark region denotes the statistical uncertainties of the method, while the lighter region illustrates the systematic uncertainties on the absolute scale.}
\label{X_galactic}
\end{figure}

The resulting calibration curve is shown in Figure~\ref{X_galactic}. This method predicts an increase in calibration factor towards higher frequencies. Noticeable bumps are visible at 50, 60 and 70 MHz, mostly as a result of binned determination of the electronic noise contribution. This indicates that a different measurement of the electronic noise can be used to improve this analysis. This, however, requires a potentially time-consuming dedicated measurement effort. Alternatively, the current results could be smoothed before applying the calibration. 

\subsubsection{Uncertainties}
The largest source of uncertainties on $X(\nu)$ follows from the contribution of the electronic noise to $T_{sky} / T_{sys}$.  This uncertainty propagates into a systematic uncertainty of 37\% on the calibration curve. It should be noted that this is a conservative estimate of the uncertainty. The second important contribution is caused by the prediction of the Galactic radiation. The absolute scaling uncertainty of different reference maps ranges up to 20\% in temperature, while the use of different models yields a difference of 5\%. The scaling uncertainties in temperature result in uncertainties on $X(\nu)$ of 9\% and 2\%, respectively. 

The statistical uncertainties contain several aspects. Per data sample, they contain all used dipoles (up to a maximum of 96), as well as data collected over the course of several years. Therefore, all occurring weather circumstances (ranging from sub-zero up to \unit{35}{^\circ C}, from dry conditions to rain or snow) are included in the spread of the calibration curve. Measured statistical fluctuations have a standard deviation of up to 5\% over the full frequency range. Fluctuations are of the same order when analyzing all events for any single antenna, so that it can be concluded that this is mostly dominated by environmental circumstances rather than fluctuations between antennas.  An overview of all different contributing uncertainties can be found in Table~\ref{tab:calgal_uncertainties}. The total uncertainties sum up to
\begin{eqnarray}
 \frac{\sigma_{X(\nu)}}{X(\nu)} = \pm 1\% \;(\text{antenna-by-antenna}) \pm 5\% \;(\text{event-by-event})
 &\pm& 38\% \;(\text{calibration}) .
\end{eqnarray}

\begin{table}[t]
    \begin{center}
       \begin{tabular}{l l l r }
         \hline
         \hline
          Uncertainty ($\sigma$) &&& Value [\%] \\
         \hline
         \hline
         {\bf Antenna-by-antenna}  && Variations between antennas          & 1 \\
         \cline{2-4}

         && {\bf Total}  &{\bf 1 }\\
&&& \\
         \hline
         \hline
         {\bf Event-by-event}  && Environmental          & 5 \\
         \cline{2-4}

         && {\bf Total}  &{\bf 5 }\\
&&& \\
         \hline
         \hline
 {\bf Calibration}                           &  &  Choice of sky model & 2 \\
                              &                  & Absolute scaling of model & 9 \\
                              &                  & Relative scaling of model & 5 \\
                              &                  & Electronic noise & 37 \\
                               \cline{2-4}

                           && {\bf Total}  &{\bf 38 }\\

          \hline
          \hline
       \end{tabular}
    \end{center}
    \caption{Summary of the uncertainties on the calibration curve in amplitude that have to be considered for the calibration on the diffuse emission from the Galaxy.}
     \label{tab:calgal_uncertainties}
\end{table}


\begin{figure}
\centering
\includegraphics[width=0.6\textwidth]{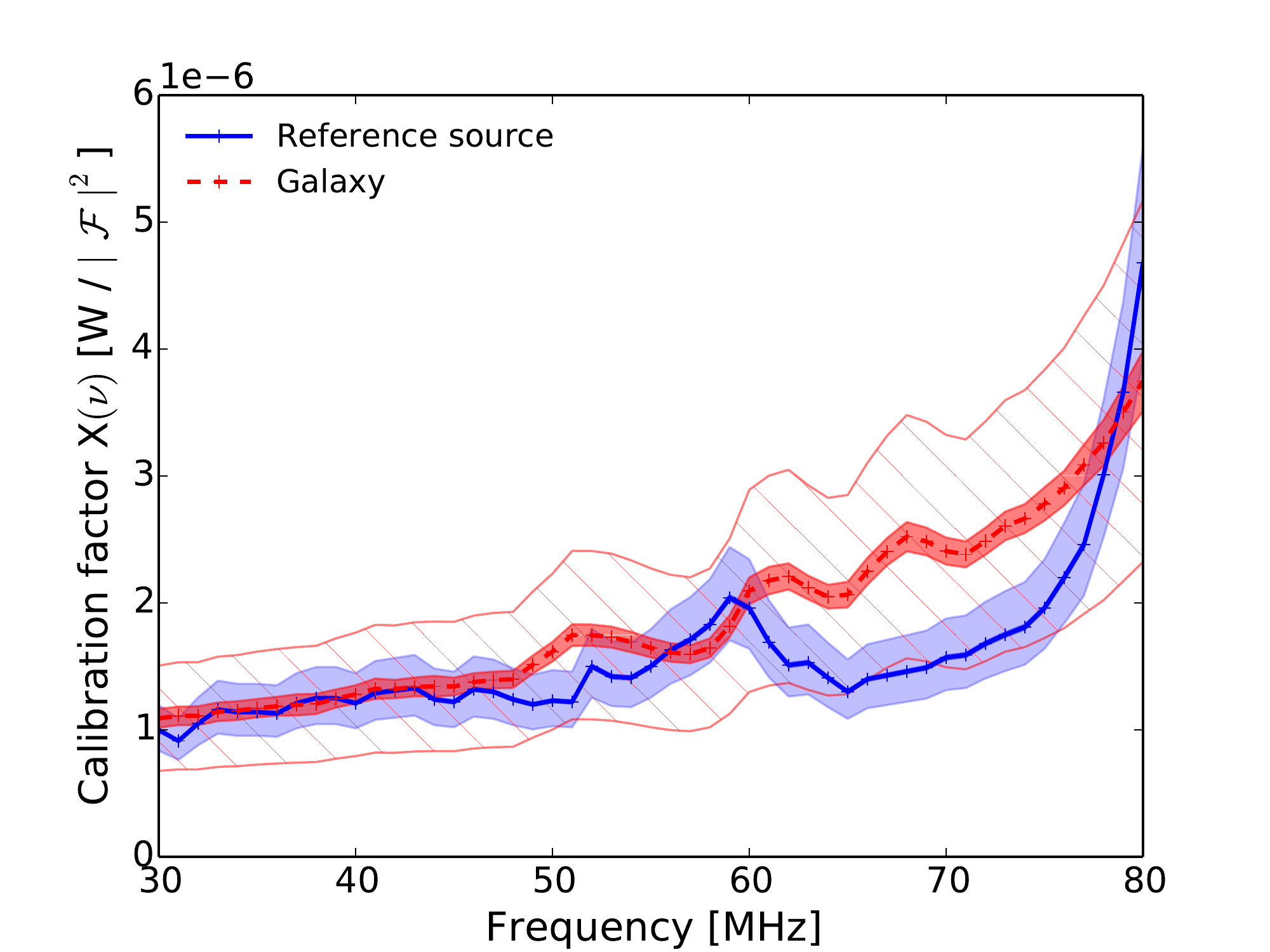}
\caption{Calibration factors as function of frequency across the LOFAR band for Galactic and reference source calibration. Both calibration curves contain statistical uncertainties of the method in the dark region, with systematic uncertainties illustrated by the lighter region (dashed for Galactic, filled for terrestrial).}
\label{bothX}
\end{figure}

\subsection{Comparison of calibration methods}
The two methods to obtain an absolute calibration are compared to each other and the consistency with other experiments as reported in the literature is checked. Since the absolute signal amplitude scales with the energy of the air shower (e.g.\ most recently \cite{GlaserICRC}), the consistency between experiments is especially relevant for studies of cosmic ray properties and the corresponding energy scale. 

\subsubsection{Comparison of Galactic and reference source calibration}
Both calibration methods result in frequency-dependent calibration factors with a very similar behavior as a function of frequency (Figure \ref{bothX}). The two curves are fully compatible below 60 MHz, and only above this frequency slight deviations are visible. Here, it is interesting to note that the shape of the two curves also deviates. This can probably be attributed to the fact that the two methods use different types of signals. While the reference source calibration exploits signals of several orders of magnitude above the noise level, the Galactic calibration relies on the noise level itself. Being essentially a simple dipole, the LBAs are mostly sensitive to the resonance frequency, meaning that for higher frequencies the antenna becomes too long (inductive) and its impedance is no longer small with respect to the LNA. Thus, the gain of the LNA decreases and the contributions of the noise budget accumulated in the coaxial cables and the several amplification stages becomes relevant. This, however, does not affect the strong signals of several orders above the noise level. Consequently, the two curves show a slightly different shape with respect to the antenna model that does not yet include the above mentioned cables and amplifiers.

Furthermore, the Galaxy is not a point-source like the reference antenna. The measurement is always an integration over all visible angles. This means that all discrepancies between antenna model and the antenna behavior with fold into the measurement. This is different for the reference source, where only the directions directly above the antenna are calibrated. The two components of the antenna model, directivity and frequency dependence, can therefore not be decoupled with the calibration on the Galactic emission. 

Both curves show a steepening slope towards the higher frequencies, which confirms the antenna calibration measurements with the drone that showed that the current antenna model describes a steeper fall-off than seen in data. Potential reasons are manifold. For example, coax cables are known to yield a signal-attenuation that increases with frequency, which has not been considered in the current model. Simplified corrections improve the situation, but this cannot fully explain the observed steepening. The obtained electronic noise corrections for the Galactic calibration result in visible bumps around 50, 60 and \unit{70}{MHz}. Finally, signal reflections and known cross-talk of closely spaced antennas can affect the averaged Galactic measurements. 

The reference source calibration is essentially only a snapshot of the data, which means that time-varying influences had to be estimated rather than measured. For the Galactic calibration, a better time-average is possible, however, the electronic noise introduces unfortunate features and currently a rather large scale uncertainty. Both results are currently completely compatible and of similar quality. 

\subsubsection{Comparison to other experiments and previous calibrations}
\begin{figure}
\centering
\includegraphics[width=0.6\textwidth]{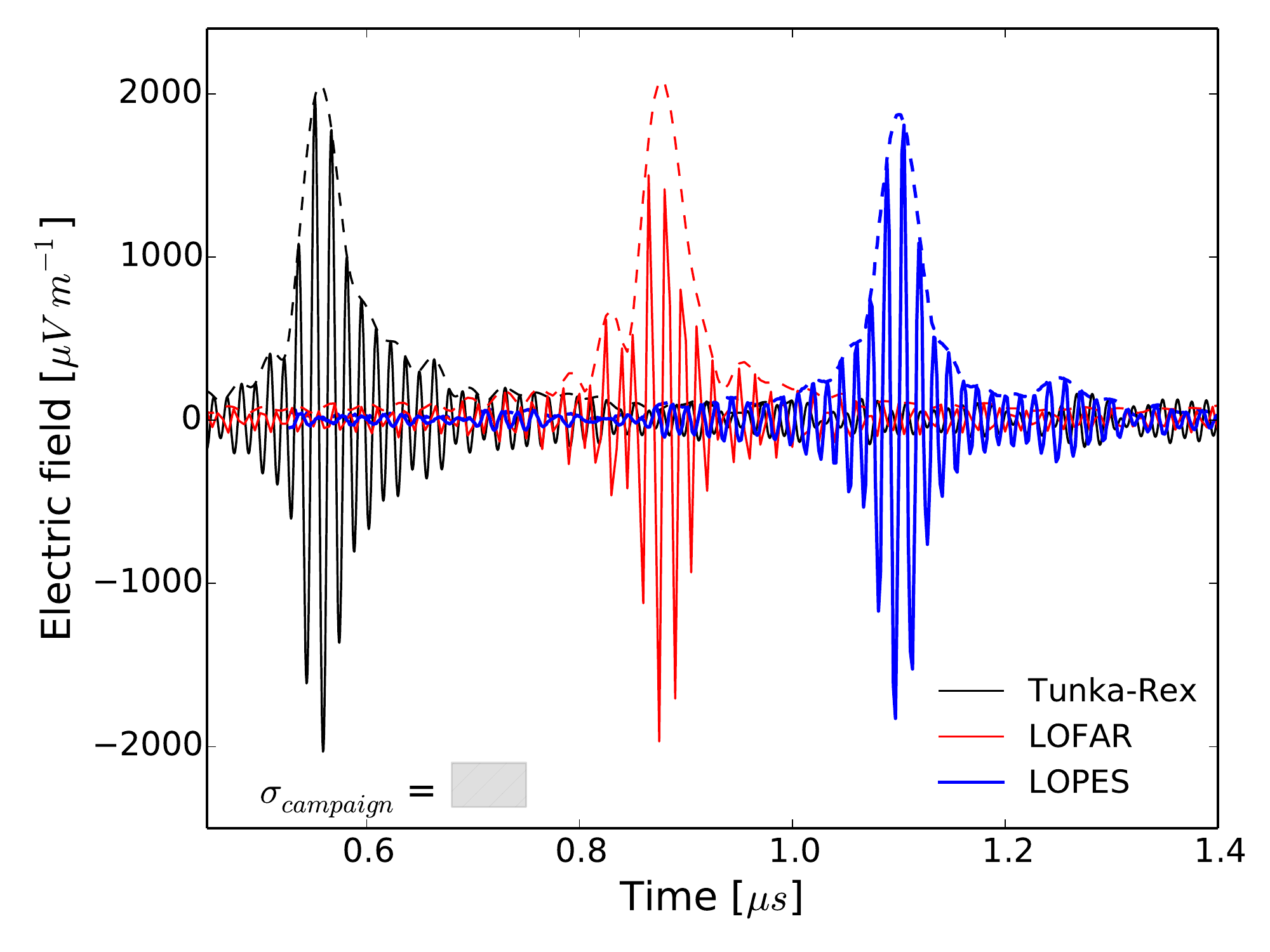}
\caption{Electric field as function of time for the three calibration measurements at (from left to right) Tunka-Rex, LOFAR and LOPES, with amplitudes scaled to match a source at 10 meter distance, up-sampled and filtered to a common band of \unit{43-74}{MHz}. The dashed lines represent the calculated Hilbert envelopes \cite{Schellart2013}  for the three pulses. They allow for a better comparison of the data that are all taken with different sampling frequencies. The size of the systematic campaign specific uncertainty on the maximum amplitude is indicated with the grey bar. These uncertainties are expected to be of the same order of magnitude for Tunka-Rex and LOPES. The pulses are offset in time from each other for a better visibility.}
\label{lofarlopestunka}
\end{figure}

A calibration using the same source antenna has been performed on radio antennas at both LOPES and Tunka-Rex. The measured voltage traces after the application of the appropriate calibration curves at the three different experiments is shown in Figure~\ref{lofarlopestunka}. As the calibration technique is designed to calibrate the detected pulses to values reported by the manufacturer of the VSQ 1000, the pulses from different experiments are required to be of the same amplitude after normalization to the same source-distance (10 m). Differences in a single measurement only occur due to different hardware response, such as group delays and sampling frequencies. The figure shows that the LOPES pulse is slightly lower than the ones from the other experiments. It should be noted that the LOPES pulse is an average of a large number of measurement campaigns and thereby slightly different. Since it is compatible within the campaign uncertainties, it gives an indication that the remaining systematic uncertainties between the experiments are well understood. 

It should be noted that the reference antenna has been re-calibrated between the published results in \cite{Nehls2008} and this campaign. The newest calibration has been performed in accordance with free-space circumstances, which matches the needs of air shower experiments. Differences between the old reference values (reflective ground) and the new values are typically a factor of two in amplitude \cite{LOPESnew}. Furthermore, the magnitude of the absolute uncertainty has been improved considerably. 

Having the same reference source at the three different experiments, allows us to ignore some absolute scaling uncertainty when comparing results from the different experiments in future publications. The largest contribution to the scale uncertainty as shown in Table \ref{tab:calcrane_uncertainties} are the method-specific factors. As opposed to the campaign specific ones, these uncertainties fully cancel out, when comparing air shower measurements at LOPES, LOFAR and Tunka-Rex. This means that the reconstructed energy of cosmic rays from radio emission of air showers becomes directly comparable, with small systematic uncertainties.

\section{Cross-check with CoREAS air shower simulations}
\label{sec:coreas}
The calibration values as obtained above have been compared to predictions from air shower simulations. The air shower simulations predict the pulse strength delivered per ideal antenna, given that shower axis position, energy, height of shower maximum and arrival direction are known. When folding the ideal signal strength with the specific antenna model, a direct comparison of model and data is possible \cite{Buitink2014}. 

A selection of high-quality air showers is routinely simulated with CoREAS \cite{Huege2013b} to determine the height of the shower maximum. In order to match the absolute signal strength to the (still) uncalibrated LOFAR data, a scaling factor $S$ is introduced, which compensates for the missing calibration. 

We compare the simulated pulse energy density (CoREAS and antenna model) and the measured pulse energy density. Each energy density is obtained by integrating the squared pulses over $i=11$ samples of \unit{5}{ns} each around the pulse maximum. This integrated value is corrected for the average noise level in each time-bin by subtracting its contribution from the integrated value \cite{Schellart2013}. To obtain the total energy density ($I_m$), both instrumental polarizations are added. The scaling factor $S$ is the ratio of the (uncalibrated) energy density derived from measurements ($I_m$) and the same quantity obtained from CoREAS simulations in combination with the antenna model, ($I_e$):

\begin{equation}
S = \frac{I_m}{I_e} \;\;\;\;\;\; [S] =  \sum_i\frac{|\mathcal{F}_i|^{2}}{J} .
\end{equation}

In future work, we plan to implement the two calibration curves in the data analysis pipeline and to perform an in-depth study of the agreement of air shower simulations with data, including the timing, polarization and frequency behavior. Here, as a first step, we compare the scaling factors $S$ obtained from every method. 

\begin{figure}
\centering
\includegraphics[width=0.6\textwidth]{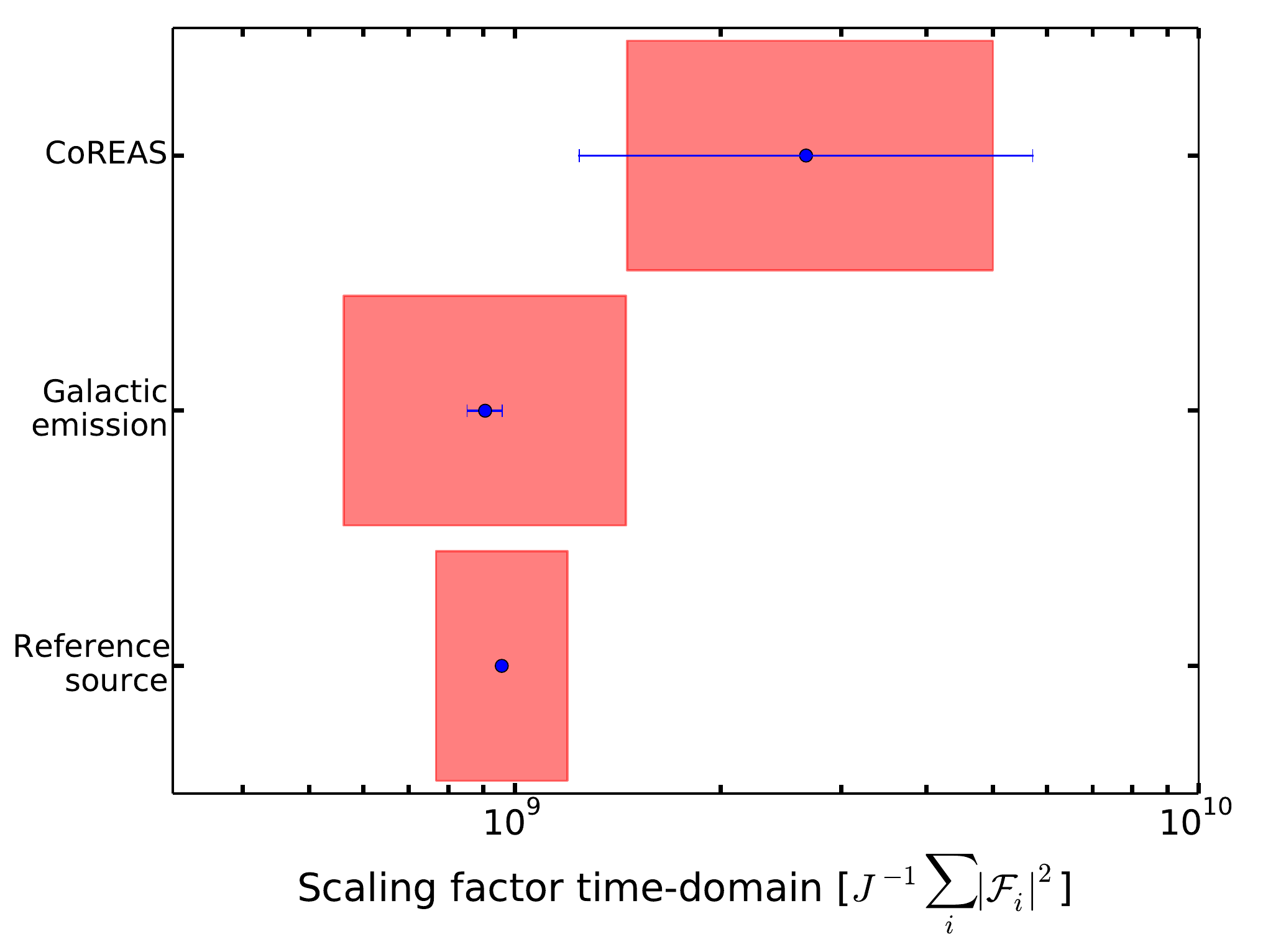}
\caption{Scaling factor $S$ as obtained in the time domain for the air shower simulations (CoREAS) and the two calibration methods (Galaxy and Reference source). Depicted values and uncertainties are the average and its standard deviation obtained from different events. The colored bands correspond to systematic uncertainties on the used method as discussed in the corresponding sections.}
\label{CoREAS}
\end{figure}

As shown in \cite{Buitink2014}, the energy used as input to the air shower simulations has to be correct to obtain a stable scaling factor. As the signal amplitude scales directly with the energy of the shower, the accuracy of the scaling factor $S$ between LOFAR data and CoREAS simulations is severely determined by the energy resolution and energy scale of the particle detectors. Furthermore, the method used to combine the fitting of particle and radio data determines the scaling factor.

The factor $S$ describing the same conversion of integrated powers has been obtained for both the reference source calibration and the Galactic calibration. In the case of the reference source, measured pulses\footnote{A frequency comb corresponds to pulses in the time-domain} are squared, integrated in the time-domain and compared to the expectations. It was found that subtracting the noise-level contributes less than 1\% in cases of rather strong signals like the ones from the reference source. For the Galactic calibration, emission is not restricted to pulses, and the entire trace is integrated and normalized to the same time-width. As power is conserved in Fourier transforms, the powers for the Galactic contribution are integrated in the frequency-domain. 

We would like to stress that the method by which $S$ is obtained varies significantly from the calibration method used earlier in this analysis to obtain $X(\nu)$. An important advantage of $X(\nu)$ is that it allows for a discussion of frequency-dependent characteristics, which the scalar value $S$ does not. $X(\nu)$ is therefore the desired quantity. Air shower pulses, however, are usually shorter than $\unit{60}{ns}$, which does not allow us to determine $X(\nu)$ with a high frequency-resolution. Furthermore, it is expected that air shower pulses show a dependency of frequency as function of distance to the shower axis and the height of the shower maximum. These effect have yet to be experimentally confirmed. Consequently, we cannot derive a third independent $X(\nu)$, but have chosen to use $S$, which is independent of frequency. 

The resulting values for $S$ for all three methods are depicted in Figure~\ref{CoREAS}, including their statistical and systematic uncertainties. The statistical uncertainties of the scaling value $S$ for CoREAS are determined by the energy resolution of the air shower measurements. Systematic uncertainties include the absolute scale of the energy of the measured cosmic rays, and uncertainties due to the air shower simulations of CoREAS. For the method currently used, the uncertainty on the absolute energy scale is estimated conservatively at $\sim$50\% \cite{Buitink2014},\cite{Thoudam2015}. The precise uncertainties due to emission models or hadronic interaction models are more difficult to establish. They can be estimated by the comparison of the two air shower simulations CoREAS and ZHAires \cite{Alvarez-Muniz2011} and how they describe the LOFAR data. The results differ by less than 30\% in $S$ for a small test-sample of showers \cite{workshop}. It cannot be established whether the differences are due to the underlying particle air shower simulations or the implementation of the calculation of the radio emission. The difference, however, provides an estimate of the uncertainty.

Both, Galactic and reference source calibration result in very similar values for $S$, which was expected due to the agreement shown in Figure~\ref{bothX}. We have to note that there seems to be some tension between CoREAS and the calibration values. However, the observed differences are of the order of magnitude of the systematic uncertainties, so the findings are still in agreement with a correct prediction of the air shower emission strength with CoREAS.  A more detailed comparison is needed to draw firm conclusions about the compatibility. A comparison also including the polarization of the signal is currently being studied, but is beyond the scope of this paper.

\section{Conclusion and Outlook}
We have presented three different methods that have been used to calibrate the low-band antennas of the LOFAR radio telescope. The first method provides detailed information about the antenna response. By positioning a reference source at different positions around an antenna, its directionality as well as the frequency response have been be measured and compared to models of the hardware. The antenna model for the LOFAR LBAs has shown to be in reasonable agreement with the measurements. It provides the foundation for secondary corrections that have been derived during the absolute calibration performed in this analysis. 

Based on this antenna model, the two other presented calibration methods have been used for an absolute calibration. Both methods, a calibration on a reference source and one on the diffuse Galactic emission, deliver results that are in good agreement with each other. The reference source method is subject to a systematic scale uncertainty of 19\%, while the Galactic calibration is dominated by the uncertainty of the noise contribution of the system of 38\% in amplitude. 

Both measurement campaigns demonstrated that a calibration with reference sources requires considerable experimental effort and has to be repeated several times to reduce systematic dependencies on weather and temperature conditions. Due to the calibration of the reference sources, the largest systematic uncertainties can hardly be reduced. Reference sources allow for very flexible testing of different components of the system and provide valuable information about the model of the antenna. They are an essential tool to understand whether the response of the hardware has been modeled correctly in software. Better calibration measurements in anechoic chambers are only possible in very large chambers at these low frequencies and will not include ground effects. It is therefore likely that field measurements will remain an important tool.  

Calibrations on astronomical sources or the diffuse emission of the Galaxy are less time-consuming. If the experiment is located in an environment with little external noise and if the antenna is designed to be sky noise limited, the data to perform this calibration can be gathered during regular operations for every single antenna of the experiment. The largest systematic uncertainty at LOFAR is the missing detailed knowledge of the contribution of the electronic noise. Dedicated measurements can reduce these uncertainties within a manageable amount of effort. This will bring the systematic uncertainty down to the uncertainty of the sky models, which are better than 10\% in amplitude and make the Galactic calibration compatible in accuracy to the reference source. 

Further improving the antenna model itself, will also help to reduce the uncertainties of the calibration curves. Using the campaign data and a new antenna model, the calibration curve can be updated without additional experimental effort. Every calibration curve is only valid in combination with the antenna model. It seems worthwhile to consider to also measure the phase response of the LOFAR system, which is included in the antenna model and especially relevant for broad-band transient data. 

The calibration curve obtained from the reference source will be used in future studies as an absolute reference scale to further understand the radio emission of air showers. Using this scale makes the results of LOFAR directly comparable to those from other experiments calibrated with the same source since the largest systematic uncertainties will not be relevant in the comparison.

Despite a desirable improvement, the current limiting uncertainty of 19\% is already competitive with respect to air shower physics. The amplitude of air shower pulses scales with the energy of the shower. If dependent only on the radio signal, the amplitude calibration will therefore determine the energy scale uncertainty. The best current energy scale uncertainty is delivered by Fluorescence telescopes and is quoted at 14\% \cite{Auger}. Particle arrays have to be calibrated against other methods, for example air shower simulations, and are therefore usually subject to systematic uncertainties larger than 19\%. Their decreased sensitivity to the chemical composition of the cosmic rays adds another systematic uncertainty. 

At LOFAR, the systematic uncertainty from the energy scale of the particle detector array limits conclusions about the correctness of the models of the radio emission of air showers. From our analysis, one could argue that there is some tension between the calibration measurements and the values derived from the air shower emission as predicted by CoREAS. However, systematic uncertainties are substantial, and no significant disagreement was found. A more dedicated comparison of air shower models and measurements, also taking the signal timing, frequency response and polarization into account, will shed more light on this issue and is currently underway. 

The calibration curve for air shower measurements can easily be applied to other measurements with the Transient-Buffer Boards at LOFAR to obtain, for example, an absolute flux scale for pulsars. Collecting suitable data to test this effort is currently ongoing.

\acknowledgments
We would like to thank the LOPES, Pierre Auger and Tunka-Rex Collaborations for sharing their experience and calibration methods. 

The LOFAR cosmic ray key science project acknowledges funding from an Advanced Grant of the European Research Council (FP/2007-2013) / ERC Grant Agreement n. 227610. The project has also received funding from the European Research Council (ERC) under the European Union's Horizon 2020 research and innovation programme (grant agreement No 640130).   We furthermore acknowledge financial support from FOM, (FOM-project 12PR304) and NWO (VENI grant 639-041-130). AN is supported by the DFG (research fellowship NE 2031/1-1).  

The calibration methods used were developed under partial support by the 'Helmholtz Alliance for Astroparticle Physics HAP' and the Networking Fund of the Helmholtz Association, Germany. 

LOFAR, the Low Frequency Array designed and constructed by ASTRON, has facilities in several countries, that are owned by various parties (each with their own funding sources), and that are collectively operated by the International LOFAR Telescope (ILT) foundation under a joint scientific policy.

\bibliographystyle{JHEP}
\bibliography{antenna_calibration}

\end{document}